%% file: main.tex
\documentclass[%
 reprint,
 amsmath,amssymb,
 aps,
]{revtex4-2}

\usepackage{graphicx}
\usepackage{subfigure}
\usepackage{dcolumn}
\usepackage{bm}
\usepackage[utf8]{inputenc} 
\usepackage[T1]{fontenc}    
\usepackage{hyperref}       
\usepackage{url}            
\usepackage{booktabs}       
\usepackage{amsfonts}       
\usepackage{nicefrac}       
\usepackage{microtype}      
\usepackage{xcolor}         
\usepackage{amsmath}
\usepackage{graphicx}
\usepackage{subfigure}
\usepackage{booktabs}
\usepackage{comment}
\usepackage[capitalise]{cleveref}

\usepackage[printonlyused]{acronym}
\input{acronyms}

\newcommand{\glitch}[1]{{\texttt{#1}}}
\newcommand{\nonconf}[1]{{\texttt{#1}}}
\newcommand{\changed}[1]{\textcolor{black}{#1}}

\begin{document}
\preprint{APS/123-QED}

\title{Classification uncertainty for transient gravitational-wave noise artefacts with optimised conformal prediction}

\author{Ann-Kristin Malz}
 \email{ann-kristin.malz@ligo.org}
\author{Gregory Ashton}
\author{Nicolo Colombo}
\affiliation{Royal Holloway University of London}

\date{\today}

\begin{abstract}
With the increasing use of \ac{ML} algorithms in scientific research comes the need for reliable uncertainty quantification. When taking a measurement it is not enough to provide the result, we also have to declare how confident we are in the measurement. This is also true when the results are obtained from a \ac{ML} algorithm, and arguably more so since the internal workings of \ac{ML} algorithms are often less transparent compared to traditional statistical methods. Additionally, many \ac{ML} algorithms do not provide uncertainty estimates and auxiliary algorithms must be applied. \Ac{CP} is a framework to provide such uncertainty quantifications for \ac{ML} point-predictors. In this paper, we explore the use and properties of \ac{CP} applied in the context of glitch classification in gravitational wave astronomy. Specifically, we demonstrate the application of \ac{CP} to the Gravity Spy glitch classification algorithm. \ac{CP} makes use of a score function, a nonconformity measure, to convert an algorithm's heuristic notion of uncertainty to a rigorous uncertainty. We use the application on Gravity Spy to explore the performance of different nonconformity measures and optimise them for our application. Our results show that the optimal nonconformity measure depends on the specific application, as well as the metric used to quantify the performance. 
\end{abstract}

\maketitle

\section{Introduction} \label{sec:intro}
\input{intro}

\section{Conformal Prediction} \label{sec:cp}
\input{cp}

\input{application}

\section{Optimising nonconformity measures}  \label{sec:nonconformity_measures}
\input{nonconformitymeasures}

\input{metrics}

\input{optimising}
\input{nonconf_comparisons}
\input{optimising_indevidual}

\section{Discussion and conclusion} \label{sec:discussion}
\input{discussion}

\begin{acknowledgments}
We want to thank Christopher Berry, Zoheyr Doctor, Ryan Fisher, Siddharth Soni, and Michael Zevin from the Gravity Spy team for their help using the Gravity Spy code and datasets and for early discussions of this work. 

This material is based upon work supported by NSF’s LIGO Laboratory which is a major facility fully funded by the National Science Foundation. 

The authors are grateful for computational resources provided by the LIGO Laboratory and supported by National Science Foundation Grants PHY-0757058 and PHY-0823459.

\end{acknowledgments}



\input{references.bbl}
\end{document}

%% file: acronyms.tex
\newacro{GW}{Gravitational Wave}
\newacro{CP}{Conformal Prediction}
\newacro{SNR}{signal-to-noise ratio}
\newacro{FAR}{false alarm rate}
\newacro{CDF}{cumulative distribution function}
\newacro{TP}{true positive}
\newacro{FP}{false positive}
\newacro{FN}{false negative}
\newacro{TN}{true negative}
\newacro{ML}{Machine Learning}
\newacro{CNN}{Convolutional Neural Network}
\newacro{ROC}{receiver operating characteristic}


%% file: intro.tex
With the first detection of gravitational waves in 2015 \cite{abbott2016GWobservation}, sourced by a pair of stellar-mass black holes colliding in another galaxy, a new way to explore the universe and a new field of astrophysics research has opened. Gravitational waves are observed using laser interferometers, such as LIGO (Laser Interferometer Gravitational-Wave Observatory) \cite{aasi2015advancedligo}, Virgo \cite{acernese2014advancedvirgo} and KAGRA (Kamioka Gravitational Wave Detector) \cite{aso2013kagra}, sensitive to relative differences in the arms of the interferometers of less than 1 part in $10^{-21}$ caused by a passing gravitational wave. We have now observed around 100 such signals \cite{abbott2023openDataO3}, but they are rare events: lasting a few seconds in year-long observing runs. 

\changed{The sensitivity of the detectors is determined by background noise; which on short timescales can be approximated as quasi-stationary coloured Gaussian noise in addition to non-Gaussian transient noise artefacts known as "glitches" \cite{abbott2020guideLIGOVirgoNoise}.} Glitches are troublesome as they often have unknown physical origins (environmental or instrumental) or are difficult to mitigate in the detectors \cite{davis2022detectorCharacterisationMitigation, abbott2020guideLIGOVirgoNoise}. They can be mistaken for gravitational wave signals, reduce the significance of signal candidates, or bias the astrophysical parameter estimation results when occurring in temporal proximity to a signal \cite{davis2022detectorCharacterisationMitigation, pankow2018MitigationGW170817}. Additionally, glitches occur at a rate of approximately 1 per minute \cite{abbott2023gwtc3} while we detect approximately 1 signal per week \cite{abbott2020prospects}, depending on the thresholds used. Thus, to improve the detection of gravitational waves and the scientific research of astrophysical events, the causes of these non-astrophysical noise artefacts must be identified and minimised in the detectors, or, alternatively, the glitches must be mitigated in the data \cite{abbott2023gwtc3, davis2022subtractingGlitchesO3}. 

We generally expect to observe astrophysical signals in all detectors observing with the required sensitivity, while glitches are caused locally only. However, the high glitch rate  \cite{davis2022detectorCharacterisationMitigation} implies that accidental coincidence between detectors is possible. Furthermore, the detectors also independently record signal-free auxiliary data which measure different aspects of the detector components and environment \cite{davis2021ligoDetectorCharacterizationO2O3, soni2024ligoDetCharO4, nguyen2021environmentalNoise}. The auxiliary channels can thus witness disturbances and can be used to help distinguish astrophysical signals from noise, as well as correlate a glitch in the strain data with noise from auxiliary sensors \cite{davis2021ligoDetectorCharacterizationO2O3, colgan2020efficientGWGlitchIdentification}. 

Glitches come in a variety of different morphologies, with each class of glitches sharing similar features \cite{davis2021ligoDetectorCharacterizationO2O3, glanzer2023dataQualityGravitySpy}. Accurate classification helps correlate glitch classes with auxiliary channels and subsequent identification of the underlying cause \cite{davis2022detectorCharacterisationMitigation}. Furthermore, identifying glitches with similar morphologies allows for prioritisation by characteristics and quantity. Thus, classifying glitches correctly is an important first step for \changed{their} mitigation, and hence is directly related to improved scientific results. 

Even when the cause of a particular glitch class cannot be identified in the detectors, correct classifications can help model the glitches and hence subtract them from the data (see for example \changed{\cite{bondarescu2023antiglitch, lopez2022simulatingtransientnoise, fernandes2023convolutional}}). The classification of glitches depending on their features is a typical \ac{ML} problem and is addressed by the Gravity Spy project for LIGO \cite{zevin2017gravityspy} and GWitchHunters for Virgo \cite{razzano2023gwitchhunters}.

Gravity Spy \cite{zevin2017gravityspy} is a citizen-science and \ac{ML} project to classify glitches in gravitational wave data. The ML algorithm consists of a \ac{CNN} \cite{wu2024advancingGravitySpy} that is trained on human-classified time-frequency-energy plots (so-called omega scans or Q-transforms \cite{chatterji2004multiresolution}) of glitches. Both citizen volunteers and the trained \ac{ML} algorithm then provide classifications for new glitches. 

\begin{figure*}
    \centering
    \subfigure[\glitch{Blip}]{\includegraphics[width=0.32\textwidth]{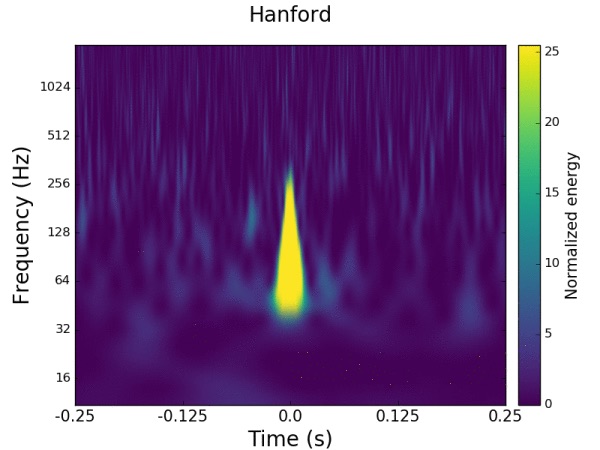}}
    \subfigure[\glitch{Tomte}]{\includegraphics[width=0.32\textwidth]{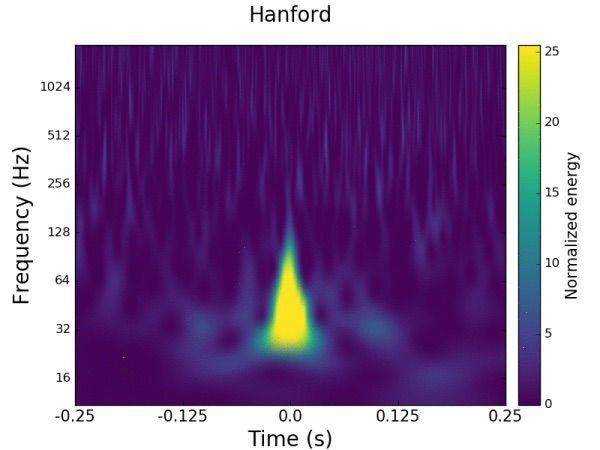}}
    \subfigure[\glitch{Koi Fish}]{\includegraphics[width=0.32\textwidth]{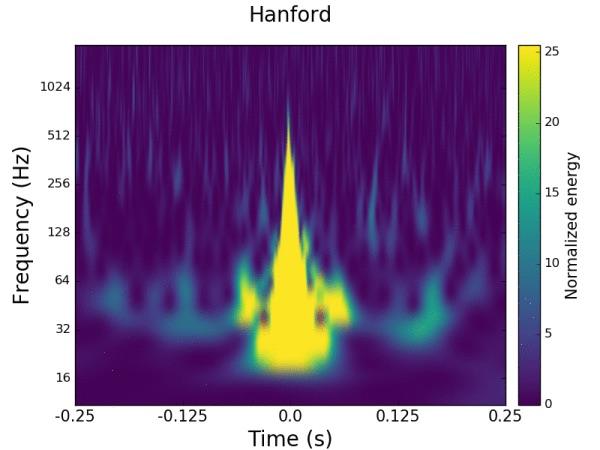}}
    \caption{Example plots of three different glitches, represented as time-frequency q-scans, as observed by the LIGO Hanford detector. The plots are the `golden images' from the Gravity Spy project, which are used to demonstrate what each glitch class looks like \cite{zevin2017gravityspy}.} 
    \label{fig:glitch_qscans}
\end{figure*}

The plots in \cref{fig:glitch_qscans} show examples of the omega scan of three different glitch classes named after their morphological features. This showcases the difficulty in correctly classifying each glitch, and the need to quantify the uncertainty of the classification, as the classes appear very similar. The similarity between some glitch classes could also suggest that they are, in fact, the same kind of glitch. 

When applying the trained Gravity Spy \ac{ML} algorithm to a particular glitch, the final layer in the \ac{CNN} returns the classification score, the estimated probability, for each possible label. The predicted label is the one that received the highest score. Due to the importance of correctly classifying the glitches, it is useful also to consider the uncertainty of this classification. For example, it would be a major advantage to have a dataset of glitches and be sure that, say, 95\% are classified correctly. The classification probabilities from the \ac{CNN} can be imperfectly calibrated, as shown by the off-diagonal points in the reliability plot in Fig. 8 in the recent Gravity Spy paper \cite{wu2024advancingGravitySpy}. Thus no inherent well-calibrated uncertainty is associated with the predictions, creating an opportunity for an external algorithm to calibrate the uncertainty. \Acf{CP} \cite{vovk2005algorithmicLearning(CPbook)} is a framework, developed in the context of \ac{ML}, to provide such uncertainty quantification for any point-prediction algorithm.

\Ac{CP} converts any heuristic notion of uncertainty from an algorithm to a rigorous uncertainty estimate \cite{angelopoulos2021gentleIntroCP}. The first step to achieving this is the definition of a score function, a so-called nonconformity measure. The nonconformity measure is built on the heuristic notion of uncertainty of the underlying pre-trained \ac{ML} model, and describes how well a sample conforms to other data. Smaller scores imply a better prediction by the underlying algorithm. This function can be defined arbitrarily as long as it returns a real-valued number that is representative of the non-conformity of a sample \cite{shafer2008tutorialonCP}. Consequently, different nonconformity measures could be used for the same application and the question is which one to choose for optimal results. Furthermore, each nonconformity measure can be parameterised and the performance can change significantly with different values of the parameters. Choosing which nonconformity measure to use thus becomes an optimisation problem. The second step is to define a calibration dataset, containing correctly labelled data. This is used to calibrate the uncertainty for the underlying algorithm, and \ac{CP} can hence be seen as learning the uncertainty of the algorithm from previous outputs.

In this paper we apply \ac{CP} to Gravity Spy, demonstrating the concept and its properties, and showing why such an uncertainty framework is beneficial. We then use this application to explore how nonconformity measures can be optimised, and we discuss how their performances on glitch classifications compare under different metrics. 

We have chosen Gravity Spy as our example algorithm, as it is a well-established \ac{ML} application in the gravitational wave community. It performs the task of image classification using a \ac{CNN}, which, in the computer science literature, is a well-explored problem but lacks inherent well-calibrated uncertainty quantification. Thus Gravity Spy provides a straightforward application for proof of concept of \ac{CP} and a good case study for future development. 

Applying \ac{CP} to gravitational wave research is a new approach that has (to the best of our knowledge) only been explored in Ref. \cite{ashton2024CP}, where a comprehensive introduction is provided, and the application of \ac{CP} to binary classification in gravitational wave search pipelines (is the event a signal or noise?) is considered. In comparison, our work discusses multi-class classification, we apply it to a \ac{ML} algorithm, and we extensively discuss the optimisation problem of choosing a nonconformity measure. Optimising nonconformity measures is relatively uncommon in the \ac{CP} literature, but is considered for example in Ref. \cite{matiz2019inductiveCPforNN}. Ref. \cite{johansson2017modelagnosticNonconf} is one of few publications where different nonconformity measures have been compared under different metrics. Thus our application of \ac{CP} to Gravity Spy as well as the comprehensive discussion of optimisation and comparison of different nonconformity measures for this application is novel. 

The remainder of this paper is structured as follows. \Cref{sec:cp} defines \ac{CP} and its properties and demonstrates the application of \ac{CP} to the Gravity Spy \ac{ML} algorithm. \Cref{sec:nonconformity_measures} introduces a few different nonconformity measures from the literature, which we modify and optimise, before comparing their performance when applied to Gravity Spy using different metrics. In \cref{sec:discussion} we discuss our results in the context of the literature. The code accompanying this paper is available from the Zenodo repository \cite{paper_code}.

%% file: cp.tex
\ac{CP} was first developed by A. Gammerman, V. Vovk, and V. Vapnik in 1998 \cite{vovk2005algorithmicLearning(CPbook), gammerman1998learningByTransduction} as a framework to quantify uncertainties in the context of \ac{ML}. \ac{CP} does not affect the underlying algorithm itself but uses past predictions from the algorithm to learn the uncertainty. Thus, there are no assumptions of the underlying model or data distributions, and no priors are needed. The only assumption required is that the data must be exchangeable. This makes \ac{CP} universally applicable to any algorithm (\ac{ML} or not) that returns a point prediction. In this paper, we will discuss the application to classification algorithms only, but the methods described are also applicable to regression algorithms.

\ac{CP} can be applied to any classification algorithm, that for each data $x$, produces a label $y$. Given a labelled dataset, \ac{CP} generalises the point-prediction from the underlying algorithm to a prediction set $\Gamma^\alpha$ of possible labels, with a user-defined \textit{error rate} $\alpha\in[0,1]$, with guaranteed \textit{validity}. Validity means that for a given $\alpha$ the true label, $\hat{y}$, is included in $\Gamma^\alpha$ with probability approximately $1-\alpha$ \cite{shafer2008tutorialonCP}. \changed{Specifically}, this is known as \textit{marginal coverage} and \changed{more formally} it can be shown \cite{angelopoulos2021gentleIntroCP} that, for $N$ calibration data points
\begin{equation}
    1-\alpha\leq\textrm{Pr}(\hat{y}\in\Gamma^\alpha)\leq 1-\alpha+\frac{1}{N+1}\,.
    \label{eq:probability_truelabel_in_set}
\end{equation}
As the number of calibration data points $N$ increases, the approximate result $1-\alpha$ is recovered. 

To apply \ac{CP}, first, a nonconformity measure $A(x, y)$ is defined. This measures the heuristic uncertainty of the underlying algorithm so that smaller scores imply a better prediction, and can be tailored to the specific application \cite{shafer2008tutorialonCP}. \changed{A common example for classification is $A(x,y)=1-f_y(x)$, where $f_y(x)$ is the classification score of the algorithm for label $y$ (later, we refer to this as the baseline nonconformity measure).}

\ac{CP} is applied in two steps, calibration and testing. This requires separate calibration and test datasets, each consisting of some data points $x$ and corresponding labels $y$. In the calibration step, the nonconformity measure $A(x,y)$ is used to calculate a \textit{nonconformity score} \changed{$s_i=A(x,y)$} for each data point $x$ in the calibration dataset. Sorting the nonconformity scores $s_i$ in ascending order, the $1-\alpha$ quantile $\hat{q}$ is calculated as
\begin{equation}
    \hat{q} = s_{\lceil(N+1)(1-\alpha)\rceil}\,,
    \label{eq:qhat}
\end{equation}
where $N$ is the total number of data points in the calibration set and the ceiling function $\lceil x\rceil$ denotes the smallest integer $\geq x$. \changed{Thus the quantile $\hat{q}$ is simply the $j^{\text{th}}$ element in the list of ordered scores, with $j=\lceil(N+1)(1-\alpha)\rceil$.} 

The nonconformity measure $A(x,y)$ can, as discussed, be any function. However, as it is used to calculate scores which are then ranked, only relative values and their ranking matter. Two nonconformity measures that are monotonic transforms of each other will thus result in exactly the same outcome under \ac{CP} \cite{shafer2008tutorialonCP}. 

In the testing step, the aim is to form a prediction set $\Gamma^\alpha$ for a test data point $x'$. Nonconformity scores are calculated for all \changed{possible} labels and the labels $y$ with scores less than $\hat{q}$ are included in the prediction set. The set of predicted labels is thus defined as
\begin{equation}
    \Gamma^\alpha = \left\{ y:A(x',y) < \hat{q} \right\}.
    \label{eq:pred_set}
\end{equation} 

\changed{For example, an image that could be either a cat or a dog is classified by an algorithm as a cat with classification scores cat: 0.8 and dog: 0.2. Using nonconformity measure $A(x',y)=1-f_y(x')$, the nonconformity scores for this example image $x'$ are $s(\text{cat})=0.2$ and $s(\text{dog})=0.8$. Assuming the quantile was previously calculated as $\hat{q}=0.35$ from some calibration data, the prediction set becomes $\Gamma^\alpha=\{\text{cat}\}$, since $0.2<0.35$, but $0.8\nless0.35$.}

Varying the error rate $\alpha$ will change the number of labels included in the prediction set, since a lower $\alpha$, and thus a higher coverage, implies that more labels will be included in $\Gamma^\alpha$ to fulfil the validity condition in \cref{eq:probability_truelabel_in_set}. As $\alpha$ goes to zero, the prediction set must include the true label with 100\% probability and will hence include all the labels. As $\alpha$ approaches one, the true label is included with 0\% probability, which implies an empty prediction set. Hence, the average number of labels included in the prediction set increases as $\alpha$ decreases. The shape of this curve depends on the problem explored as well as the chosen nonconformity measure. The value of $1-\alpha$ when the prediction set size changes from a single label to include more than one label is known as the \ac{CP} \textit{confidence} \cite{shafer2008tutorialonCP}. A discussion of alternative confidence definitions can be found in \cite{ashton2024CP}.

\changed{\subsection{Mondrian conformal prediction}}
We are also interested in \textit{conditional coverage} for each class. For example, we might want a set of data points with a specific label that is certain to be at least 95\% accurate, as would be critical, e.g. when using \ac{ML} for medical diagnoses. To guarantee validity for each class individually, Mondrian (label-conditional) \ac{CP} can be applied \cite{ding2024classConditionalCP, vovk2012conditional}. In Mondrian \ac{CP}, the data is split by class, and \ac{CP} is applied for each class separately. Thus, the conditional, and by extension the marginal, labels are guaranteed to obey \cref{eq:probability_truelabel_in_set}. The number of calibration data $N$ in \cref{eq:probability_truelabel_in_set} now becomes the number of data points per label $N_y$, thus increasing the error and hence requiring a bigger dataset to allow for small error rates $\alpha$. 

For Mondrian \ac{CP} \changed{the $1-\alpha$ quantile in \cref{eq:qhat}} becomes label conditional; the calibration step is performed separately for each class, and a different quantile \changed{$\hat{q}_y= s_{\lceil(N_y+1)(1-\alpha)\rceil}$} is obtained for each label $y$. \changed{In the testing step, the calculated nonconformity score for each possible label $y$ for a test data point $x'$ is compared to the corresponding quantile $\hat{q}_y$ and \cref{eq:pred_set} becomes $\Gamma^\alpha=\left\{y:A(x',y)<\hat{q}_y\right\}$.}

\changed{As we are interested in conditional coverage for each glitch class, we will use Mondrian \ac{CP} throughout this paper.} \\

%% file: application.tex
\subsection{Application to Gravity Spy} \label{sec:application_to_gs}
We now apply \ac{CP} to the Gravity Spy glitch classification algorithm. Applying the trained Gravity Spy \ac{ML} algorithm to a particular glitch outputs the most likely class label (\texttt{ml\_label}) and its classification score (\texttt{ml\_confidence}), as well as an array of the classification scores corresponding to each glitch class. These classification scores are the output of the final layer in the \ac{CNN} and are used as the probability distribution of the classifier \cite{zevin2017gravityspy}. Hence, they provide the heuristic uncertainty for the algorithm, which we can use as input to our nonconformity measures. 

\subsubsection{The dataset} \label{sec:dataset}
To apply \ac{CP}, we need a dataset of glitches which contains both the true label and the predicted label for each glitch. There are multiple Gravity Spy datasets available. For the work in this paper, the `retired' dataset, available from Ref. \cite{gravityspy_retired_dataset}, has been used as it already contains all the information we need. This dataset contains the citizen scientist and \ac{ML} classifications of glitches from the first three observing runs of the LIGO detectors. All glitches in the dataset have been classified by the \ac{ML} algorithm, as well as received at least one citizen volunteer classification \cite{zevin2024gravityspyLessonsLearned}. For each glitch, the dataset contains the \ac{ML} classification scores for all classes, the \ac{ML} predicted label (\texttt{ml\_label}), as well as the \texttt{final\_label}, which is the combined volunteer and \ac{ML} classification. Thus, all the information we need is included in the dataset, and there is no need for us to rerun the \ac{ML} algorithm.

\changed{The} `true label' of a glitch is required to calibrate the algorithm. Since there is no ground truth, human-only classifications would be the next obvious choice. However, there were no human-classified glitch datasets available to us that contained enough glitches of each class to calibrate \ac{CP}. Thus, we have chosen to use the \texttt{final\_label} in the `retired' dataset \cite{gravityspy_retired_dataset} as the `true label', which combines the human and \ac{ML} classifications scores \cite{zevin2017gravityspy}. Our `true label' might thus not always be accurate and, hence, the apparent performance of the algorithm according to the results in this dataset does not match the (significantly better) performance reported in Ref. \cite{zevin2024gravityspyLessonsLearned}. Furthermore, the Gravity Spy \ac{ML} algorithm \cite{glanzer2023dataQualityO3GravitySpy} our dataset \cite{gravityspy_retired_dataset} is based upon has since been improved, see Ref. \cite{wu2024advancingGravitySpy}. Nevertheless, our goal is to provide a proof of concept of how to apply \ac{CP}.

The distribution of glitches within this dataset is far from uniform, for example, there are 327262 \glitch{Scattered Light} and 1167 \glitch{Wandering Line} glitches. We want to identify a set where there are enough glitches in each class to apply \ac{CP}. Hence, we randomly choose 1500 glitches of each class (or all glitches of classes where fewer are available) to make up our dataset. 

\subsubsection{Application}
\changed{First}, we use the output array of classification scores to determine a simple nonconformity measure
\begin{equation}
    A(x,y)=1-f_y(x)\,,
    \label{eq:nonconf_measure_simple}
\end{equation}
where $f_y$ is the classification score for label $y$ as given by Gravity Spy. The nonconformity measure defined in \cref{eq:nonconf_measure_simple} is the most common for classification problems and is sometimes referred to as the hinge loss, see e.g. Ref. \cite{johansson2017modelagnosticNonconf}. In the remainder of this paper, we will refer to it as the \nonconf{baseline} measure. However, this choice is not necessarily optimal. In \cref{sec:nonconformity_measures}, we will discuss and compare alternative measures and introduce approaches to parameterise and optimise them.

Next, we use the curated Gravity Spy dataset, split \changed{equally} into a calibration and test set, and follow the calibration steps as described above, to obtain a quantile $\hat{q}_y$ for each glitch class. We can then apply \ac{CP} to the Gravity Spy output for a random test glitch (our test data point $x'$); \changed{calculating the nonconformity scores for each possible label and including all labels $y$ with a nonconformity score smaller than $\hat{q}_y$ in the prediction set $\Gamma^\alpha$.} 

\changed{To demonstrate the application, we consider the following examples where we calculate the quantiles $\hat{q}_y$ using two different choices of error rate: $\alpha=0.32$ (68\% probability that the true label is included in the prediction set) and $\alpha=0.1$ (90\%). We then create prediction sets for a test glitch for each of the $\alpha$ values. Using a \glitch{Blip} glitch (randomly chosen from the test dataset) that was classified correctly by Gravity Spy with a classification score of $0.998$ gives the following results:}

\begin{verbatim}
    68%: Blip -> {Blip(0.998)},
    90%: Blip -> {Blip(0.998), Tomte(0.0001)},
\end{verbatim}
where the glitch class on the left is the true label and \{.\} represents the prediction set. The numbers in parenthesis represent the classification score from Gravity Spy for that glitch class.

As another example, we pick a case where Gravity Spy classifies the glitch incorrectly. Here, a \glitch{Tomte} glitch is classified as a \glitch{Koi Fish} with a classification score of $0.49$. Applying \ac{CP} gives prediction sets: 
\begin{verbatim}
    68%: Tomte -> {Koi_Fish(0.49), Tomte(0.35)},
    90%: Tomte -> {Blip(0.0003), Koi_Fish(0.49), 
                   Tomte(0.35)}.
\end{verbatim}
In this example, Gravity Spy makes an incorrect classification but \ac{CP} still includes the correct label in the prediction set, and is guaranteed to do so \changed{approximately} 9/10 times for $\alpha=0.1$.

\begin{figure}
    \centering
    \includegraphics[width=0.5\textwidth]{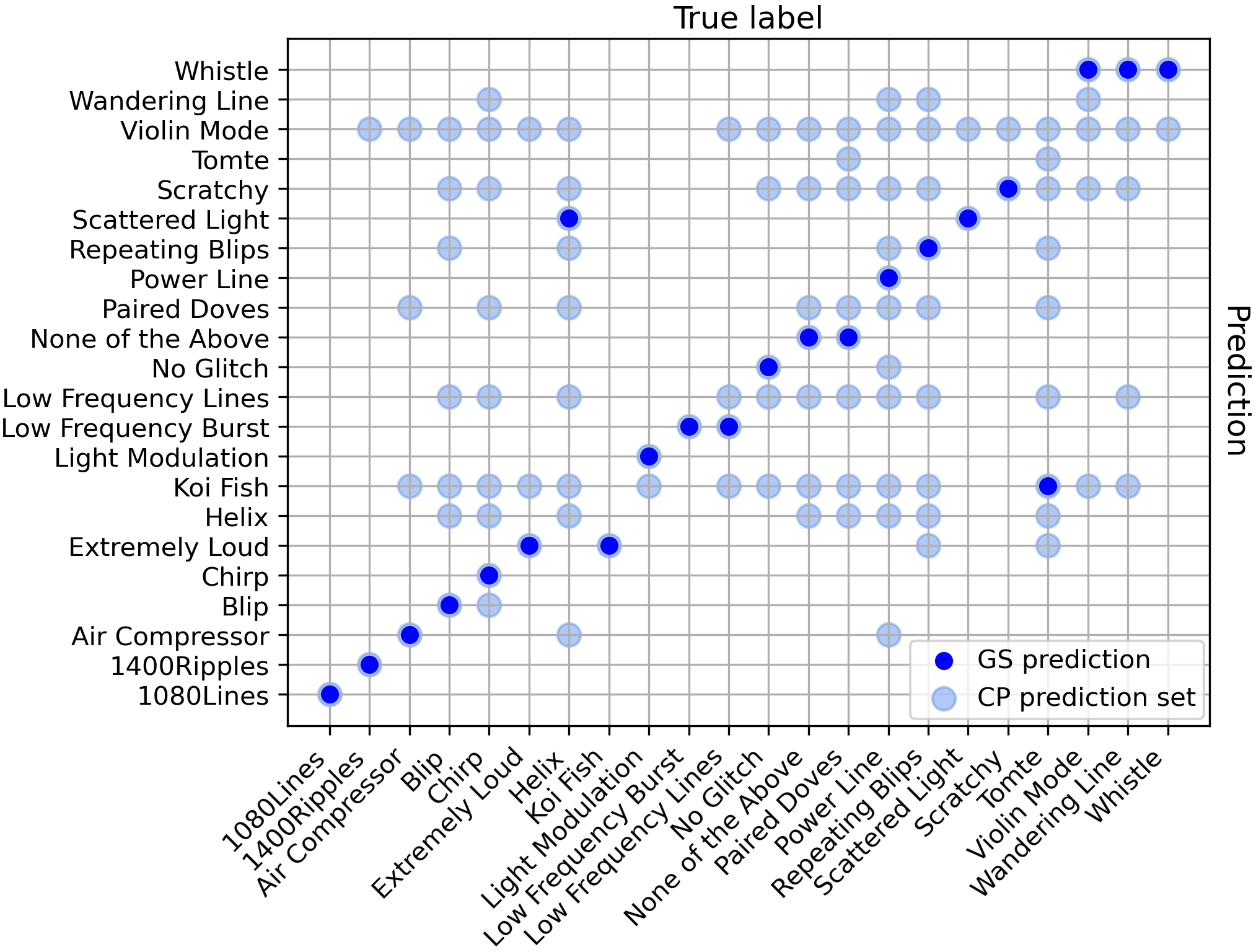}
    \caption{Scatter plot for Gravity Spy (GS) predictions on a few chosen test glitches (dark blue points). Points on the diagonal represent correct predictions, as the true label on the x-axis and the predicted label on the y-axis agree. The plot also contains the \changed{\ac{CP} set} (using $\alpha=0.1$) for each example glitch, represented as bigger light blue points. For each true label, there are one or several labels included in the prediction set.} 
    \label{fig:cp_gs_scatterplot}
\end{figure}

\begin{figure}
    \centering
    \includegraphics[width=0.48\textwidth]{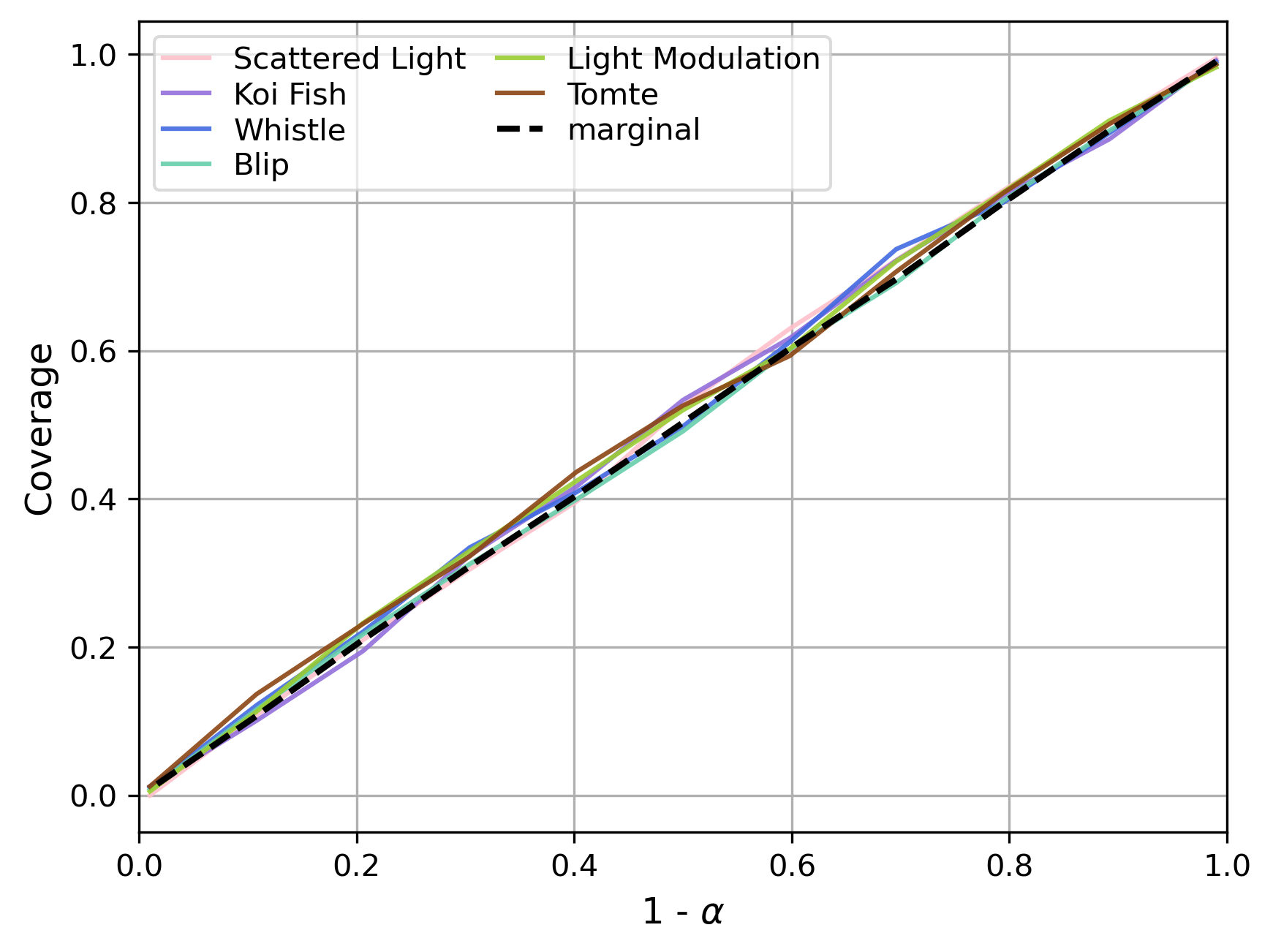}
    \caption{\ac{CP} conditional and marginal coverage, $\textrm{Pr}(\hat{y}\in\Gamma^\alpha)$, for varying error rate $\alpha$. The coloured lines represent the conditional cases for \changed{a few} different glitch classes, and the black dashed line represents the marginal case. This plot \changed{illustrates} the \ac{CP} validity property from \cref{eq:probability_truelabel_in_set}.} 
    \label{fig:validity}
\end{figure}

Extending the example to multiple test glitches, we now consider one test data point from each glitch class (randomly chosen from our test dataset) and show the results of applying \ac{CP} (with $\alpha=0.1$) in \cref{fig:cp_gs_scatterplot}. The figure shows one example for each glitch class to demonstrate \ac{CP} and is not indicative of the general behaviour of the respective glitch classes. \changed{For each test glitch on the $x$-axis, the predicted label by Gravity Spy, as well as all labels included in the prediction set, are shown on the $y$-axis.} Correct predictions by Gravity Spy are shown on the diagonal, where the predicted label matches the true label. We observe that the \changed{\ac{CP} set} varies in size, and the true label is included in the set for all but two of the test glitches shown in the plot. This \changed{illustrates} the validity property, as the true label is only guaranteed to be in the prediction set 90\% of the time. The plot in \cref{fig:cp_gs_scatterplot} shows the varying prediction set size and thus the varying uncertainty of each prediction. Furthermore, some glitch classes, such as \glitch{Violin Mode} and \glitch{Koi Fish}, are often incorrectly included in the prediction set, implying that they are more likely to be confused with other glitch classes in this specific example. 
To demonstrate the concept of validity from \cref{eq:probability_truelabel_in_set} on a larger test set, we calculate the marginal and label-conditional coverage for varying $\alpha$ \changed{(choosing a few representative example glitches for the conditional cases)}. The result is shown in \cref{fig:validity}, where the diagonal line confirms the statement of validity. This is similar in nature to the Probability-Probability (P-P) plots \cite{thode2002testingNormality}, which test the agreement of datasets or if a model fits the data, and are commonly used to verify the performance of various parameter estimation algorithms. It is also similar to reliability plots \cite{brocker2007increasingReliability}, which can be used to evaluate probabilistic predictions.

\begin{figure}
    \centering
    \includegraphics[width=0.48\textwidth]{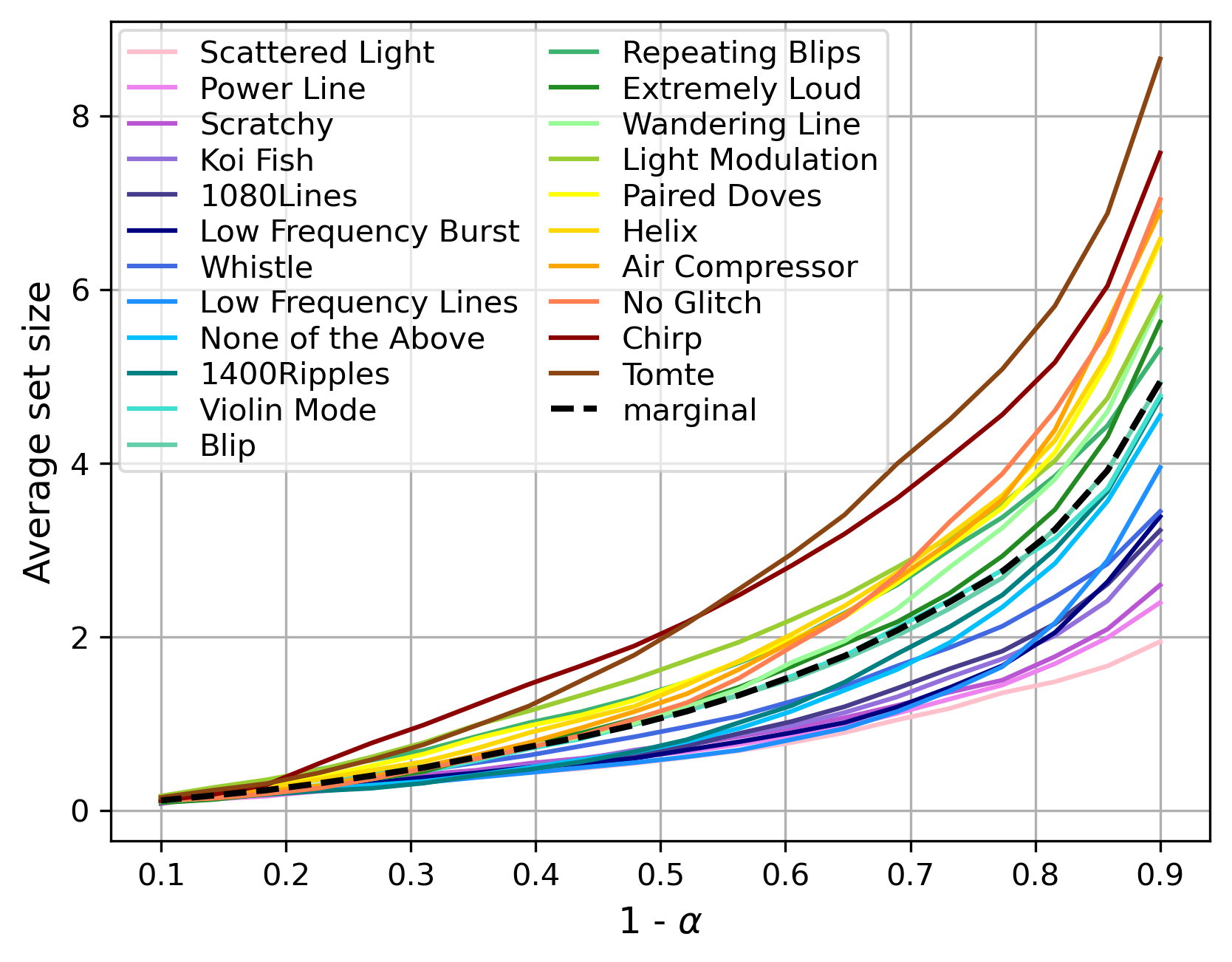}
    \caption{Average number of labels in the \changed{\ac{CP} set} $\Gamma^\alpha$, for varying error rate $\alpha$. The coloured lines represent the conditional cases for each glitch class (defined by the true labels) and the black dashed line represents the marginal case.} 
    \label{fig:average_nr_labels}
\end{figure}

We note that the marginal coverage is on the diagonal within the expected Poisson error due to the finite sample size,  while the label-conditional lines deviate. This is explained by the smaller datasets giving a larger Poisson counting error and is included by the last term in the validity guarantee in \cref{eq:probability_truelabel_in_set}, where the smaller label-conditional datasets (fewer data points $N$) give a larger interval and thus are allowed to deviate further from the diagonal in \cref{fig:validity}.
Finally, we investigate how the average prediction set size varies with the chosen error rate. One might be tempted to choose a low error rate $\alpha$ to get a high probability of the true label being included in the prediction set, but the cost of a lower error rate is a larger uncertainty in the form of a larger prediction set, as shown in \cref{fig:average_nr_labels}. The average number of labels can also be computed for each class separately, thus showing how some glitch classes are easier to classify and have higher average \ac{CP} confidence than others. \changed{The confidence can be read from the plot by investigating at what $1-\alpha$ value each line reaches an average set size of two.} For example, from \cref{fig:average_nr_labels} we can see that a \glitch{Tomte} glitch \changed{(the brown, upper-most line)} has significantly lower confidence (50\%) than a \glitch{Scattered Light} glitch \changed{(the pink, bottom-most line)} with confidence 90\%. This illustrates how the similarity of, for example, \glitch{Tomte} and \glitch{Blip} glitches (see \cref{fig:glitch_qscans}) often results in large prediction sets compared to the more easily uniquely classified \glitch{Scattered Light} glitch. The plot in \cref{fig:average_nr_labels} is cut off at $1-\alpha=0.9$ for clarity, as the set size is equal to the total number of glitches (22) when $\alpha=0$. 
Having demonstrated how to apply \ac{CP}, we can now discuss the benefits. For example, if a certain glitch has a prediction set of size one, we can guarantee that this label is the true label with probability $1-\alpha$. Another objective might be to create a set of glitches of a certain class, with known uncertainty. For example, we can collect a set of glitches that have all been classified as \glitch{Tomte}, apply \ac{CP} and determine the \ac{CP} confidence. We could then choose to include only those glitches that are above a certain confidence threshold, say $1-\alpha=0.9$, thus creating a set where each glitch is guaranteed to be a \glitch{Tomte} with $90\%$ certainty.

%% file: nonconformitymeasures.tex
There are many different nonconformity measures in the literature that could be applied instead of the \nonconf{baseline} measure in \cref{eq:nonconf_measure_simple}. In this section, we review common examples and explore their performances on our Gravity Spy dataset. Furthermore, to find the optimal versions of each nonconformity measure for our application, we parameterise each function to then be optimised. An overview of the parameterised nonconformity measures is given in \cref{tab:nonconformity_measures}, where $\beta$, $\gamma$ and $\nu$ are the \changed{tunable parameters}. Note that we have generalised several of the nonconformity measures compared to the papers referenced, \changed{by adding additional tunable parameters}. 

\input{nonconformitymeasures_table}

\subsection{Nonconformity measures}
All of the nonconformity measures are based solely on the classification scores $f_y$ output from Gravity Spy for each label $y$, as this encodes the heuristic uncertainty which \ac{CP} transforms into a rigorous one \cite{angelopoulos2021gentleIntroCP}. However, if additional information were available this could also be used to inform the nonconformity measure. The first four nonconformity measures in \cref{tab:nonconformity_measures} are adaptions on the \nonconf{baseline}, making use of the softmax function, adding weighting in the form of a cross-entropy term and a combination of both. Meanwhile, the \nonconf{maxscore2} measure uses the largest and second-largest classification scores as a weight to the \nonconf{baseline} measure. 

The \nonconf{margin2} measure makes use of the largest classification score, excluding the score of the label $y$ currently considered. The parameter $\gamma\geq0$ makes the measure sensitive to small changes in $f_y$, since decreasing $\gamma$ increases the importance of $f_y$ compared to the scores of the other labels \cite{papadopoulos2007CPwithNN}. 

The \nonconf{CNN} measure was developed specifically for use on a convolutional neural network and is, like the \nonconf{margin2} measure, constructed from the classification score of the considered label, $f_y$, and the largest score when excluding $f_y$. The parameter $\gamma$ provides a trade-off between the two terms \cite{matiz2019inductiveCPforNN}.

The \nonconf{Brier} measure includes the classification scores of all possible labels so that the final nonconformity score is affected even if there is minor confusion about the true label \cite{johansson2017modelagnosticNonconf}. The normalisation factor $\mathcal{Y}$ represents the total number of labels.

It is worth noting that most of the nonconformity measures reduce to the \nonconf{baseline}, \cref{eq:nonconf_measure_simple}, for specific choices of values for \changed{the tunable} parameters $\beta$, $\gamma$ and $\nu$.

%% file: nonconformitymeasures_table.tex
\begin{table*}
  \caption{Nonconformity measures}
  \label{tab:nonconformity_measures}
  \centering
  \begin{ruledtabular}
  \begin{tabular}{llll}
    Name     & Definition $A(x,y)$ & Reference\\
    \midrule
    Baseline & $1-f_y(x)$  & \cite{vovk1999machine}    \\
    Softmax  & $1-f_{\beta y}, \quad f_{\beta y} = [{\rm softmax}(\beta f)]_y = \frac{e^{\beta f_y}}{\sum_{y'} e^{\beta f_{y'}}}$ & \cite{luo2024entropy}\\
    Entropy  & $(1-f_y(x))|1 + \gamma h(f)|^{\nu}, \quad h(f) = -\sum_{y'} f_{y'} \log(f_{y'})$ & \cite{luo2024entropy}     \\
    \changed{Entropy softmax} & $(1-f_y(x))|1+\gamma h(f_\beta)|^{\nu}$ & \cite{luo2024entropy}  \\
    Margin2 & $\max_{y' \neq y}(f_{y'}) |f_y+\gamma|^{|\nu|}$, \quad $\gamma\geq0$ & \cite{papadopoulos2007CPwithNN} \\
    Maxscore2 & $\frac{1-f_y(x)}{|1+\gamma_1 f_{\mathrm{max}} + \gamma_2 f_{\mathrm{2ndmax}}|}$ & This work \\
    CNN     & $1-\gamma f_y + (1-\gamma)\max_{y'\neq y}(f_{y'})$, \quad $\gamma \in [0,1]$ & \cite{matiz2019inductiveCPforNN}   \\
    Brier  & $\frac{1}{|\mathcal{Y}|}\sum_{y'} |{\bf 1}(y' = y)- f_{y'}|^{\nu}$ & \cite{johansson2017modelagnosticNonconf}   \\
  \end{tabular}
  \end{ruledtabular}
\end{table*}

%% file: metrics.tex
\subsection{Metrics} \label{sec:metrics}
To optimise the nonconformity measures and compare them, we first need to define what we mean by `optimal' performance. This is not necessarily straightforward, as different applications of \ac{CP} have different purposes and thus varying definitions of what is optimal. In this section, we describe three different metrics that can be used for such optimisations and comparisons; the \textit{average prediction set size} \cite{johansson2017modelagnosticNonconf}, the number of correct predictions of set size one, so-called \textit{singletons} \cite{johansson2017modelagnosticNonconf} and the $F_1$ score \cite{humphrey2022f1, wu2024advancingGravitySpy}. \changed{We have chosen these metrics because they are commonly used in the context of machine learning and \ac{CP} in the literature, but other metrics could also be defined and considered. Each metric has different advantages, and the choice of metric depends on what the user considers an optimal outcome.}

We define the average prediction set size as
\begin{equation}
    \texttt{set\_size}=\frac{1}{N} \sum_n^N |\Gamma^\alpha_n| \,,
    \label{eq:av_setsize}
\end{equation}
where $|.|$ defines the set size and $N$ are the number of test data. The size of the prediction set $\Gamma^\alpha$ is an inherent feature of \ac{CP} which is linked to the certainty of predictions. Minimising the average set size implies more certain predictions and can hence be seen as minimising the uncertainty. 

The average number of singletons is defined as
\begin{equation}
    \texttt{singleton} = \frac{1}{N} \sum_n^N {\bf 1}(|\Gamma^\alpha_n|=1) \,,
\end{equation}
where we make use of the same notation as in \cref{eq:av_setsize} and {\bf 1} is an indicator function. Singletons are often useful when we want to classify something uniquely. For example, when classifying a set of gravitational wave events as signals or noise, one might want to maximise the number of events that are uniquely classified as signals. In general, singletons are the metric to choose if the purity of a dataset is valued, such as in population studies. 

The $F_1$ score is \changed{the harmonic mean of precision and recall,} defined as
\begin{equation}
    F_{1}=\frac {precision \cdot recall}{precision + recall } = \frac {2\mathrm {TP} }{2\mathrm {TP} +\mathrm {FP} +\mathrm {FN}}\,,
    \label{eq:f1}
\end{equation}
where \ac{TP} is the number of data correctly predicted as "positive", \ac{FP} the number incorrectly predicted as "positive", and \ac{FN} the number incorrectly predicted as "negative". The number of data correctly predicted as "negative" are the \ac{TN}. The $F_1\in[0,1]$ score is thus defined such that a higher score implies a better prediction, as given by the trade-off between the \ac{TP} rate and the \ac{FP} and \ac{FN} rates.

The $F_1$ score is generally defined for binary classification and does not directly map to \ac{CP}, since \ac{CP} returns a prediction set rather than a single prediction. Therefore, we extend the $F_1$ score so that all classes are considered as "positive" in turn, and the $F_1$ score is calculated for each label separately. Let us demonstrate this with an example. 

\begin{itemize}
    \item Blip $\rightarrow$ \{Blip\}: TP(Blip) += 1\,,
    \item Blip $\rightarrow$ \{Blip, Tomte\}: TP(Blip) += 1\,, \\ FP(Tomte) += 1\,,
    \item Blip $\rightarrow$ \{Tomte, Koi\_Fish\}: FN(Blip) += 1\,, FP(Tomte) += 1\,, FP(Koi\_Fish) += 1\,,
    \item Blip $\rightarrow$ \{\}: FN(Blip) += 1\,,
\end{itemize}
Here, the first test data point, consisting of a true label, \glitch{Blip}, and a prediction set, \{Blip\}, gives true positive \ac{TP}(Blip) $=1$ and all other counts are zero. 

Taking all four test data points above and considering \glitch{Blip} glitches only, this set of examples has TP(Blip)=2, FP(Blip)=0, and FN(Blip)=2 thus giving 
\begin{equation}
    F_1(\texttt{Blip})=\frac{2\cdot2}{2\cdot2+0+2}=0.67\,, 
\end{equation}
when applying \cref{eq:f1}.

After calculating the $F_1$ scores per label, they can be combined by taking the average, also known as macro $F_1$ \cite{lipton2014optimalF1}, over all classes. There are several alternative ways to combine the individual scores, for example using the geometric mean, but these are not considered here.  

\subsubsection{ROC curve}
\begin{figure}[h]
    \centering
    \includegraphics[width=0.45\textwidth]{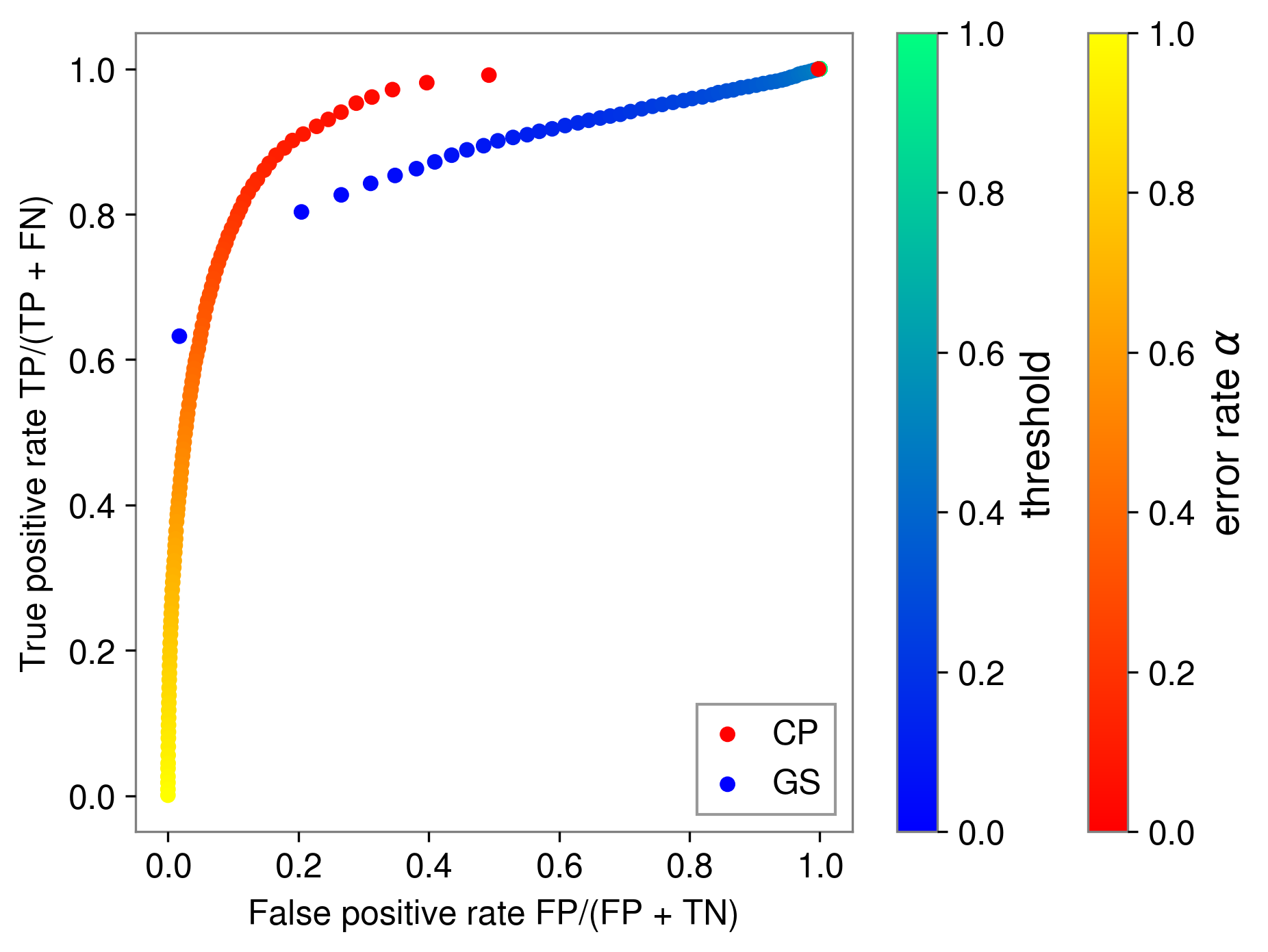}
    \caption{\ac{ROC} curves for \ac{CP} for varying error rate $\alpha$ and Gravity Spy (GS) for varying the threshold.}
    \label{fig:ROC}    
\end{figure}

The measured \ac{TP}, \ac{FP}, \ac{FN} and \ac{TN} values can also be used to determine the true positive rate, defined as TP/(TP+FN), and the false positive rate, defined by FP/(FP+TN). By varying the error rate $\alpha$, the trade-off between the true positive rate and the false positive rate can be shown with a \ac{ROC} curve \cite{cowan1998statistical}, see the red-yellow curve in \cref{fig:ROC}. Here we have used the standard definition of a \ac{ROC} curve, as defined for binary classification, with the caveat that the \ac{TP}s, \ac{FP}s, \ac{FN}s and \ac{TN}s are the values summed over all glitch classes. This demonstrates how a small error rate, which is generally desirable, achieves a high true positive rate, but at the cost of also increasing the false positive rate. The ideal case is in the top-left corner of the plot, where the true positive rate is maximised and the false positive rate minimised. 

The blue-green line in \cref{fig:ROC} is the \ac{ROC} curve for our Gravity Spy dataset, where the true positive and false positive rates are calculated for multi-class classification (an illustrative multiclass confusion matrix demonstrating this can be found for example in Fig. 3 in Ref. \cite{Betancourth2023Progressive}). We observe that for higher false positive rates, the \ac{CP} curve has higher true positive rates than the Gravity Spy curve. However, interestingly the two curves intersect and the Gravity Spy curve has a higher true positive rate at low false positive rates. Furthermore, we can compare the area under the \ac{ROC} curve (AUC), which has a value of $0.929$ for \ac{CP} and
$0.861$ for Gravity Spy. With these two results, we show that using \ac{CP} to calibrate the Gravity Spy predictions can improve the \changed{overall classification} performance. 

%% file: optimising.tex
\subsection{Optimising nonconformity measures} \label{sec:optimising_nonconf}
To optimise the nonconformity measures in \cref{tab:nonconformity_measures}, we use both a grid search and \texttt{Scipy} optimize \cite{2020SciPy} with the \texttt{L-BFGS-B} method \cite{byrd1995limitedMemoryAlgorithm,zhu1997algorithm}. From the grid search, we can obtain an understanding of the topology of the chosen parameter space, while using \changed{the \texttt{L-BFGS-B} method in} \texttt{Scipy} is more efficient to solve the optimisation problem. We perform the optimisation separately for each metric, and find the best parameters by maximising the $F_1$ score, maximising the number of singletons and minimising the average set size, respectively. Note that the prediction sets will not be affected if the nonconformity measure is changed monotonically \cite{shafer2008tutorialonCP}.

Due to the discrete nature of the underlying function, optimisation algorithms that use auto-differentiation can struggle to find the global maxima. On large scales, our metrics appear to vary smoothly with the parameters in the nonconformity measures. However, if the scale over which the parameters are varied becomes comparable to $1/N$, where $N$ is the size of the calibration dataset, it is no longer approximately smooth but becomes visibly discreet. The optimisation can be improved by increasing the step size of the algorithm, to ensure it continues on from local maxima. Alternatively, applying a differentiable approximation of the $F_1$ score could be considered if the objective is improved optimisation. For our purpose of demonstrating that nonconformity measures can be optimised, using grid search and \texttt{Scipy} is sufficiently accurate. 

For the \nonconf{entropy} and \nonconf{entropy softmax} nonconformity measures, there is degeneracy for all metrics along both axes; where one of the parameters is zero the measure reduces to the \nonconf{baseline} and we obtain unchanged scores. \changed{The degeneracy makes the optimisation difficult as no one optimum exists, and unchanged scores along an axis implies that any parameter value gives an equally good nonconformity measure so no improvements are obtained.} To avoid this, we add a regularisation, $\mathcal{R}$, in the optimisation, such that we instead maximise $F_1-\mathcal{R}$ and $\texttt{singleton}-\mathcal{R}$, and minimise $\texttt{set\_size}+\mathcal{R}$. We use $\mathcal{R}=\rho(\gamma^2+\nu^2)$ and $\mathcal{R}=\rho(\nu^2+\beta^2)$, respectively for the two nonconformity measures, where $\rho=0.00001$. Large parameter values are thus penalised and the degeneracy is removed. The small factor $\rho$ ensures that although we break the degeneracy along the axes, the regularisation does not affect the overall results.

To perform the optimisation, we first split our dataset (\changed{as described in \cref{sec:dataset}}) \changed{equally} into an optimisation and an evaluation set. The optimisation set is used to find the optimal parameters and the evaluation set is used to calculate the scores of each metric for the obtained best parameters. \changed{Each of the optimisation and evaluation datasets are then split equally into calibration and test set} to apply \ac{CP}. We repeat the optimisation five times, using different calibration-test data splits of the optimisation dataset and can thus obtain standard deviations on the optimised parameters.

In the following three subsections, we will discuss and show the optimisation results for each of the three metrics for some of our nonconformity measures. All the optimisations in these sections use an error rate of $\alpha=0.1$.

\subsubsection{Results: $F_1$ score}
\begin{figure*}
    \centering
    \subfigure[\nonconf{Entropy}]{\includegraphics[width=0.235\textwidth]{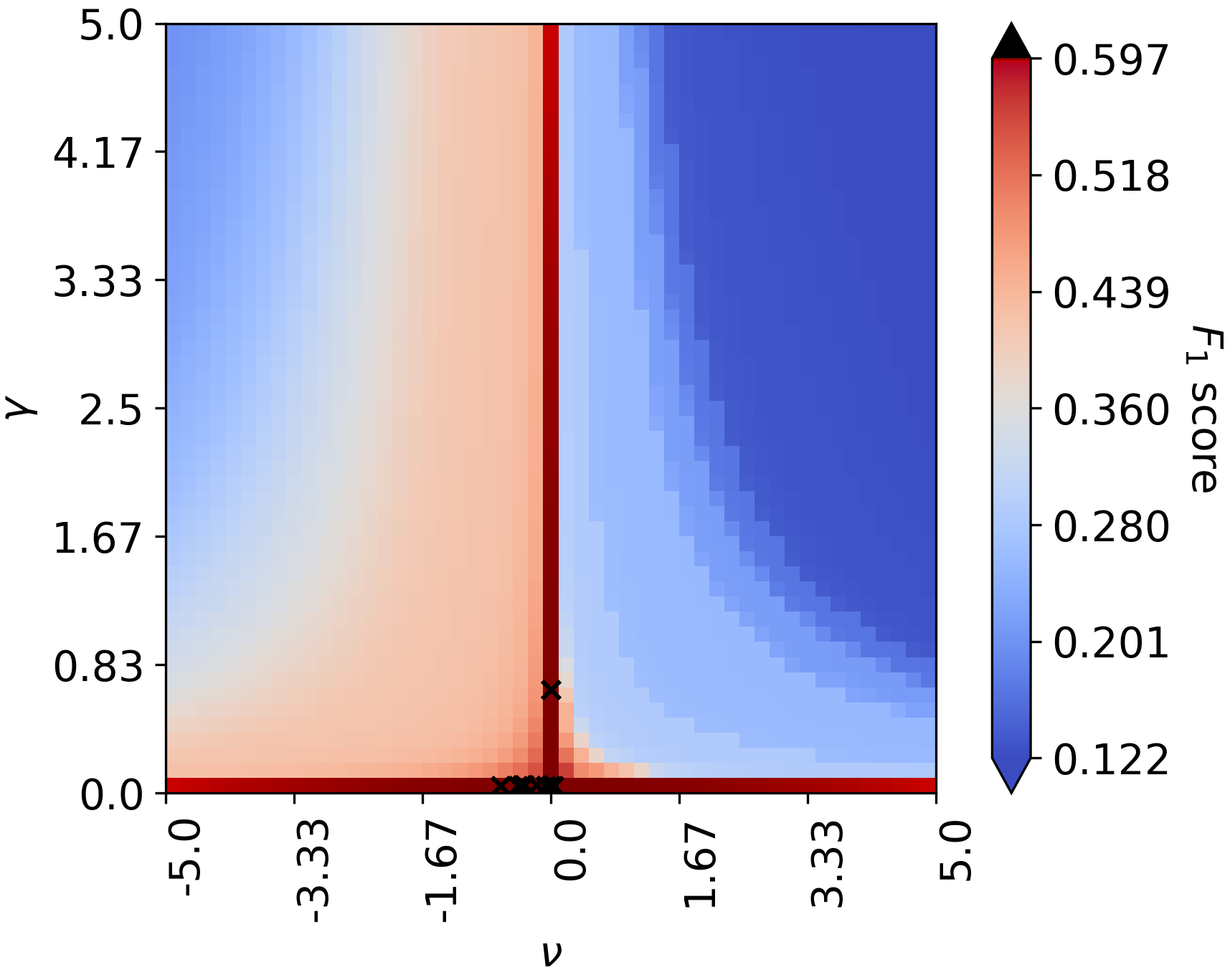} \label{fig:gridd_scipy_f1_entropy}}
    \subfigure[\nonconf{Entropy softmax}]{\includegraphics[width=0.235\textwidth]{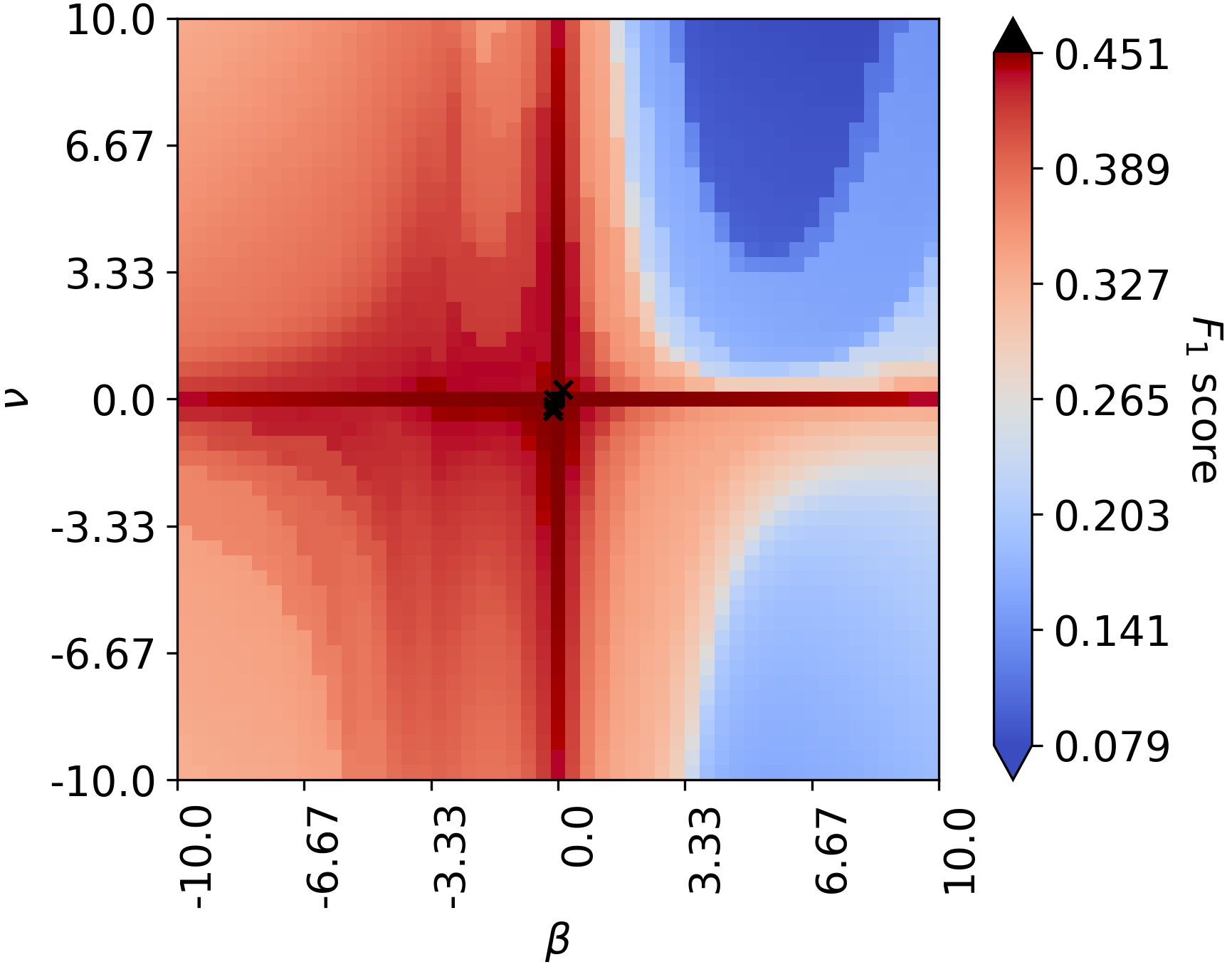} \label{fig:gridd_scipy_f1_entropy_softmax}}
    \subfigure[\nonconf{Margin2}]{\includegraphics[width=0.235\textwidth]{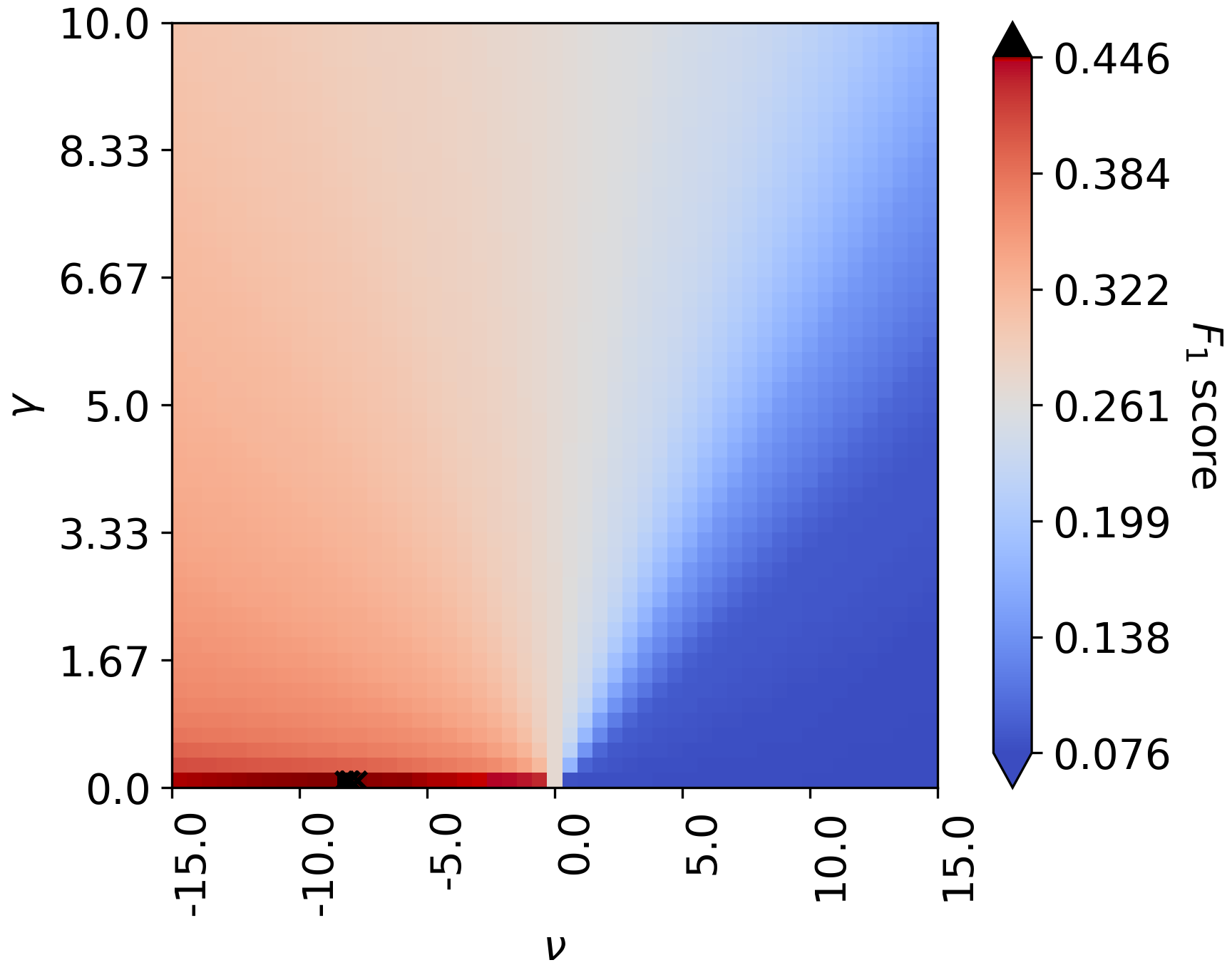} \label{fig:gridd_scipy_f1_margin2}}
    \subfigure[\nonconf{Maxscore2}]{\includegraphics[width=0.235\textwidth]{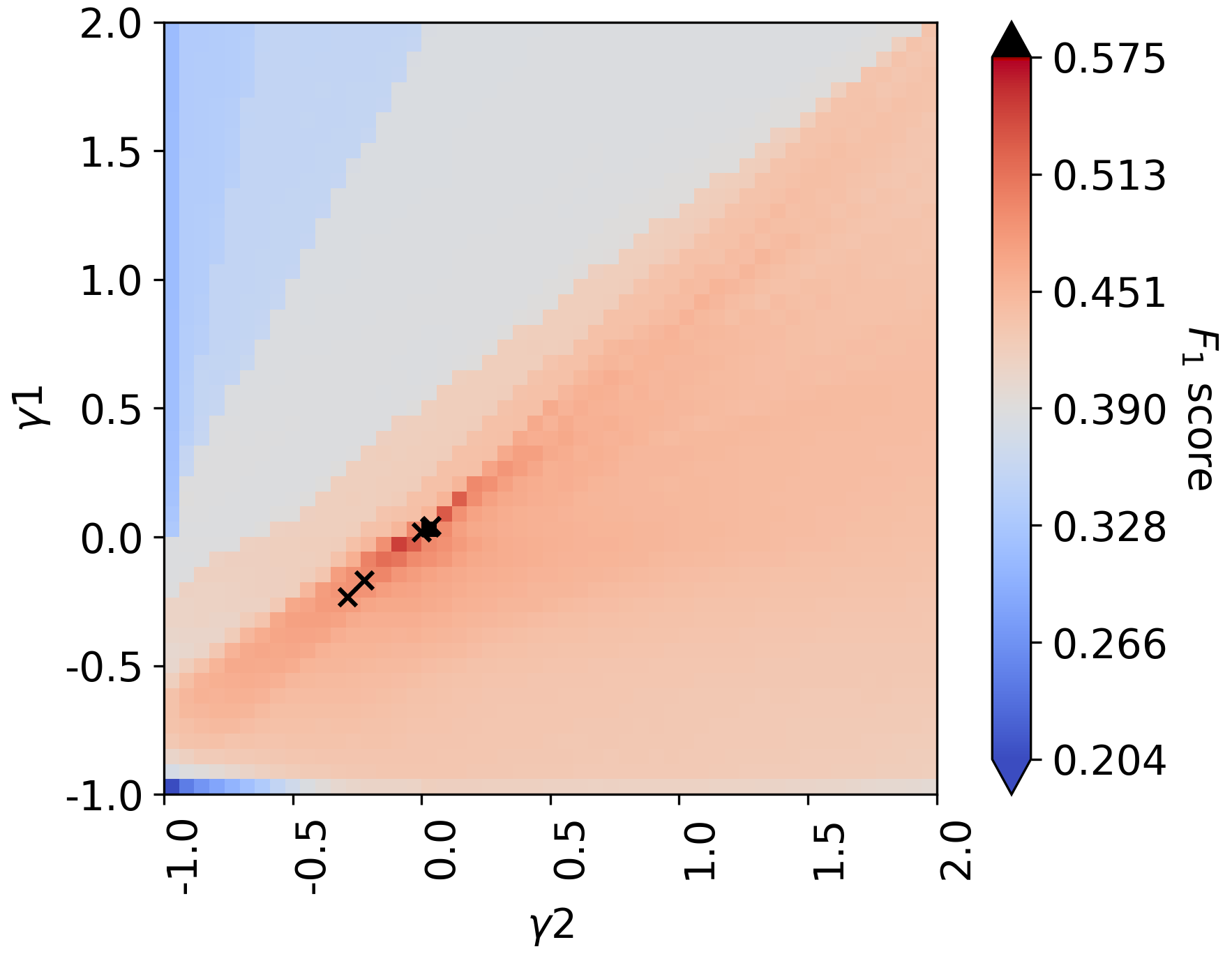} \label{fig:gridd_scipy_f1_maxscore2}}
    \caption{Grid plots with Scipy optimisations (black cross) using the $F_1$ score for four example nonconformity measures. \changed{The definition of each nonconformity measure is given in \cref{tab:nonconformity_measures}.}} 
    \label{fig:gridd_scipy}
\end{figure*}

To demonstrate the optimisation process, \cref{fig:gridd_scipy} shows \changed{the parameter space} grid plots of the $F_1$ scores as calculated for the two varying parameters in the respective nonconformity measures, overlaid with the results from several \texttt{Scipy} optimisation runs. \changed{The plots visualise the complicated topology of the parameter space for the different nonconformity measures and show that some regions of the parameter space are greatly preferred over others.}

Studying the individual nonconformity measure equations in \cref{tab:nonconformity_measures}, it is clear that, for all the measures but \nonconf{softmax}, \nonconf{margin2} and \nonconf{Brier}, certain parameter choices reduce the measure to the \nonconf{baseline}. For example, $\gamma=0$ or $\nu=0$ in the \nonconf{entropy} measure recover the \nonconf{baseline}. When applying the $F_1$ score optimisation to these nonconformity measures, they all optimise to the \nonconf{baseline}. As discussed, setting either of the parameters to zero for the \nonconf{entropy} and \nonconf{entropy softmax} measures reduces them to the \nonconf{baseline} measure and results in the highest $F_1$ scores. Applying the regularisation, this behaviour is still visible, as shown in \cref{fig:gridd_scipy_f1_entropy} and \cref{fig:gridd_scipy_f1_entropy_softmax}, but the degeneracy is removed and the optimisation improved. The \nonconf{maxscore2} nonconformity measure also optimises to the \nonconf{baseline} measure, for $\gamma_1=\gamma_2=0$, as shown in \cref{fig:gridd_scipy_f1_maxscore2}. The \nonconf{margin2} measure in \cref{fig:gridd_scipy_f1_margin2} does not reduce to the \nonconf{baseline} for any parameters. It is optimised with $\gamma=0$ and $\nu=-8.8\pm1.5$, which gives $F_1$ scores comparable to the \nonconf{baseline}, see discussion in \cref{sec:comparison}. Despite the appearance in the plot, there is no degeneracy along $\gamma=0$ when inspecting the values.

\subsubsection{Results: singletons}
\begin{figure*}
    \centering
    \subfigure[\nonconf{Entropy}]{\includegraphics[width=0.235\textwidth]{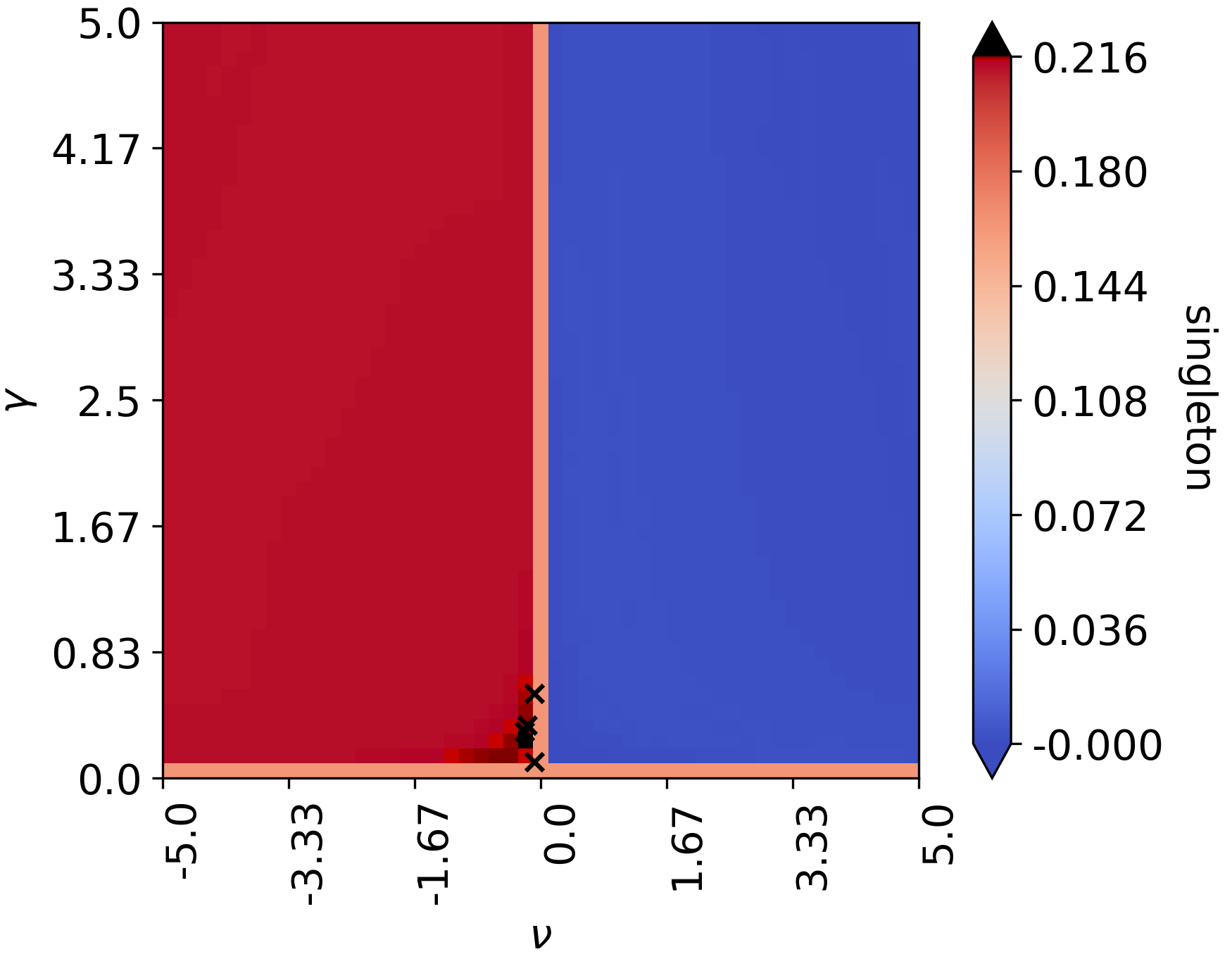} \label{fig:gridd_scipy_singleton_entropy}}
    \subfigure[\nonconf{Entropy softmax}]{\includegraphics[width=0.235\textwidth]{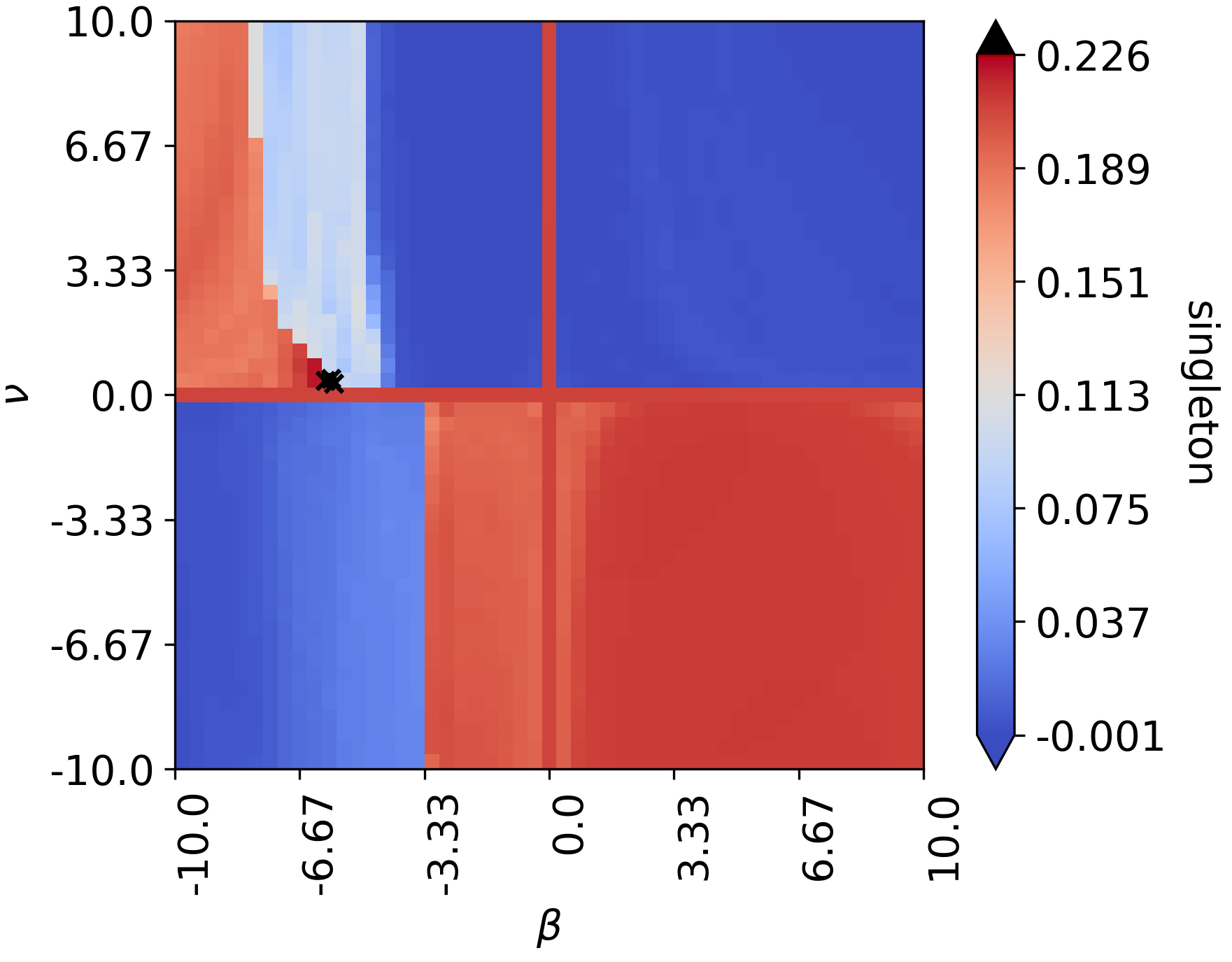} \label{fig:gridd_scipy_singleton_entropy_softmax}}
    \subfigure[\nonconf{Margin2}]{\includegraphics[width=0.235\textwidth]{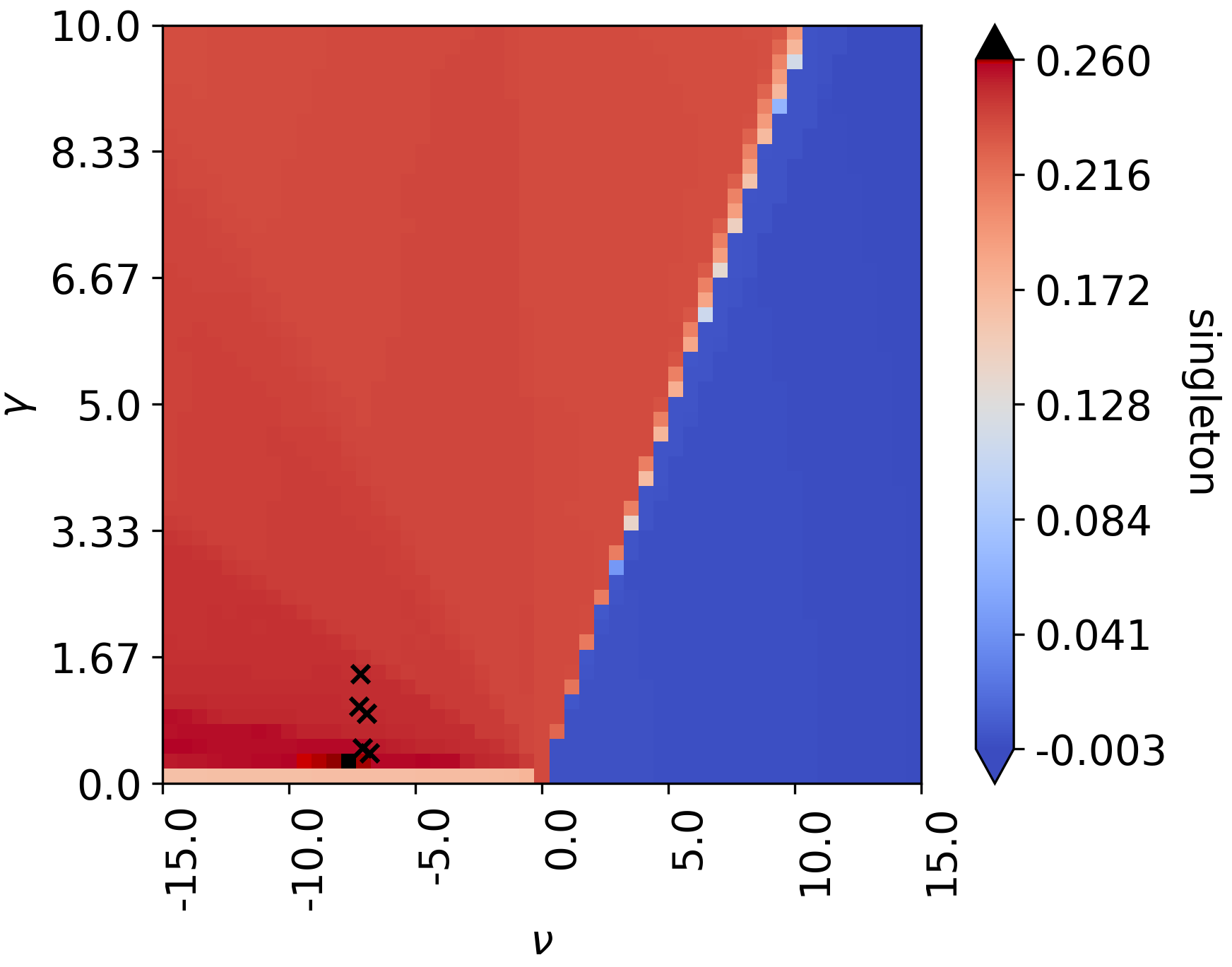}}
    \subfigure[\nonconf{Maxscore2}]{\includegraphics[width=0.235\textwidth]{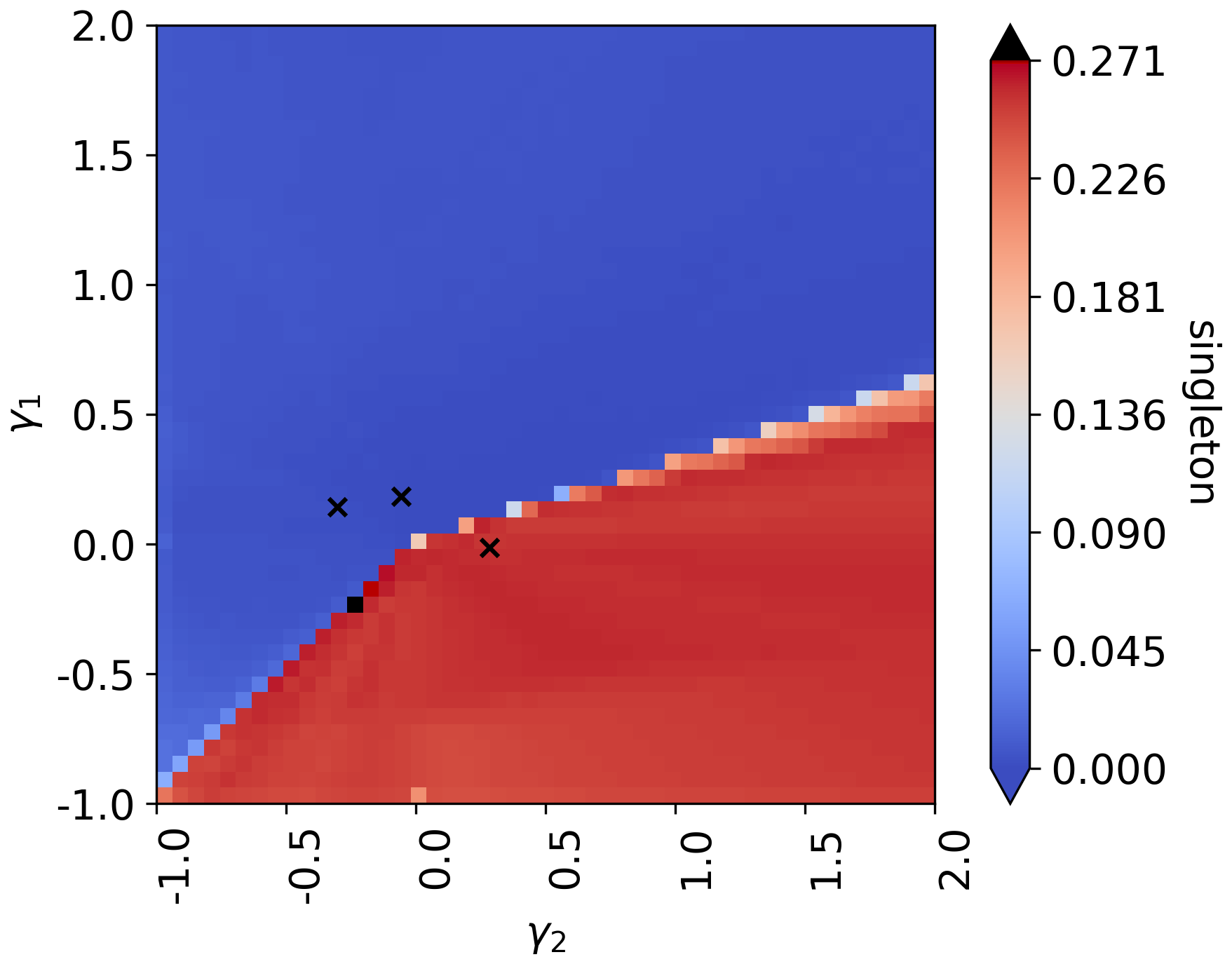} \label{fig:gridd_scipy_singleton_maxscore2}}
    \caption{Example grid plots for four different nonconformity measures with Scipy optimisations (black cross), maximising the number of singletons. \changed{The definition of each nonconformity measure is given in \cref{tab:nonconformity_measures}.}} 
    \label{fig:gridd_scipy_singletons}
\end{figure*}

When maximising the number of singletons, we observe that the nonconformity measures no longer optimise to values that reduce them to the \nonconf{baseline}, as can be seen in \cref{fig:gridd_scipy_singletons}. In fact, all nonconformity measures shown in \cref{fig:gridd_scipy_singletons} can be optimised to give better results than the \nonconf{baseline} measure when using singletons as the comparison metric. 

We also note that there is partial degeneracy for the singleton plots, however, this is mainly due to the scores being very close together and thus not distinguishable on the colour scheme of the plots. 

\subsubsection{Results: average set size}
\begin{figure*}
    \centering
    \subfigure[\nonconf{Entropy}]{\includegraphics[width=0.235\textwidth]{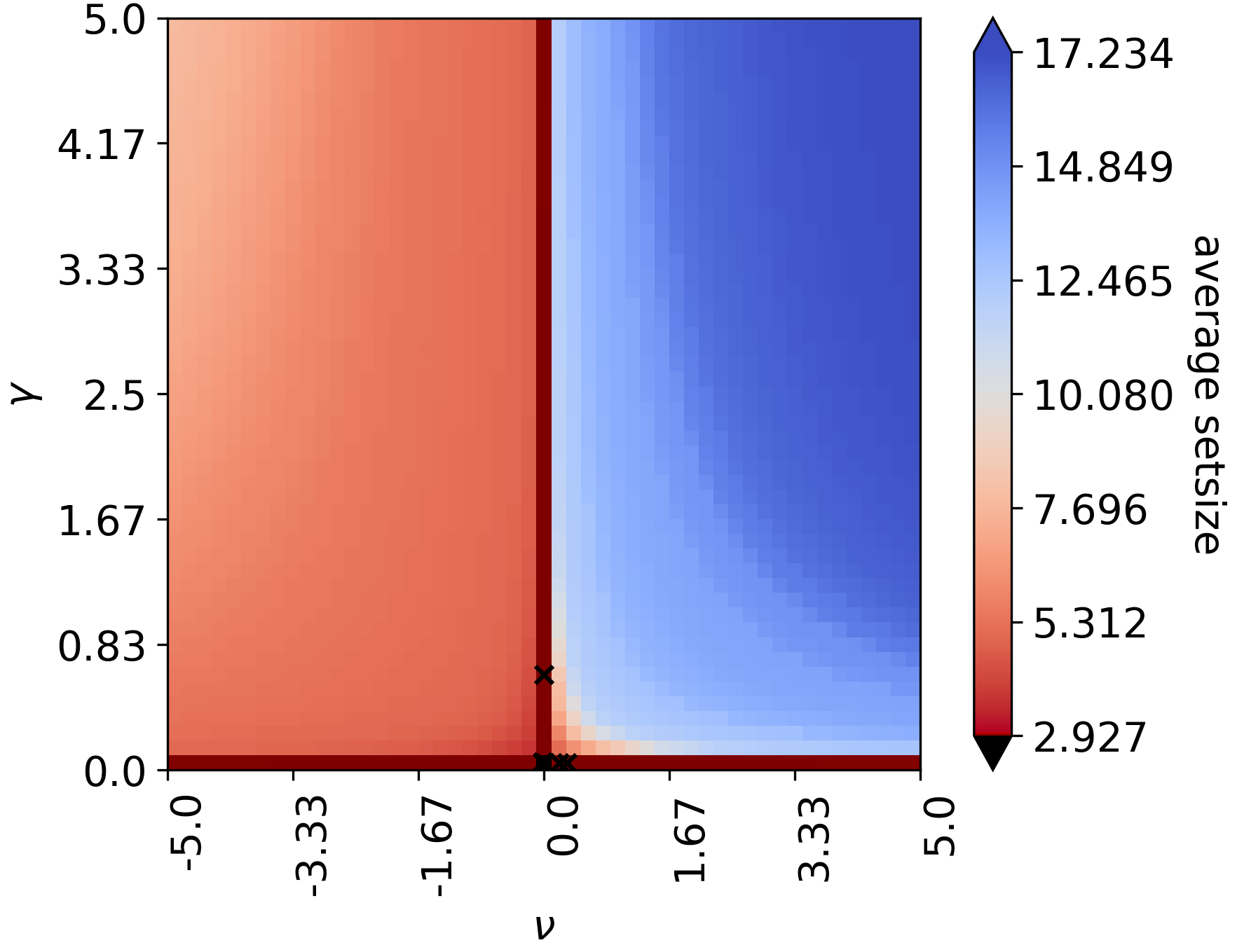} \label{fig:gridd_scipy_av_setsize_entropy}}
    \subfigure[\nonconf{Entropy softmax}]{\includegraphics[width=0.235\textwidth]{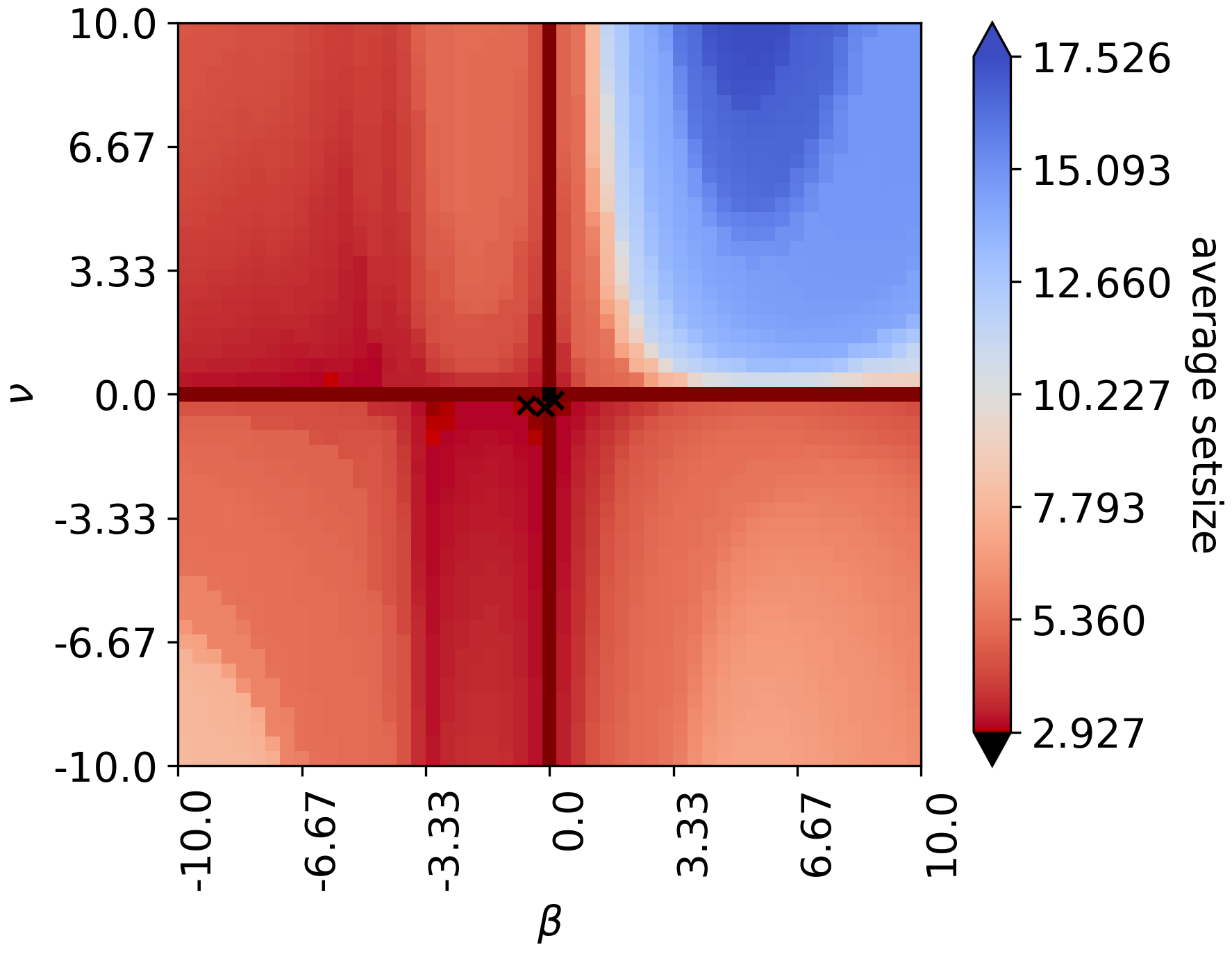} \label{fig:gridd_scipy_av_setsize_entropy_softmax}}
    \subfigure[\nonconf{Margin2}]{\includegraphics[width=0.235\textwidth]{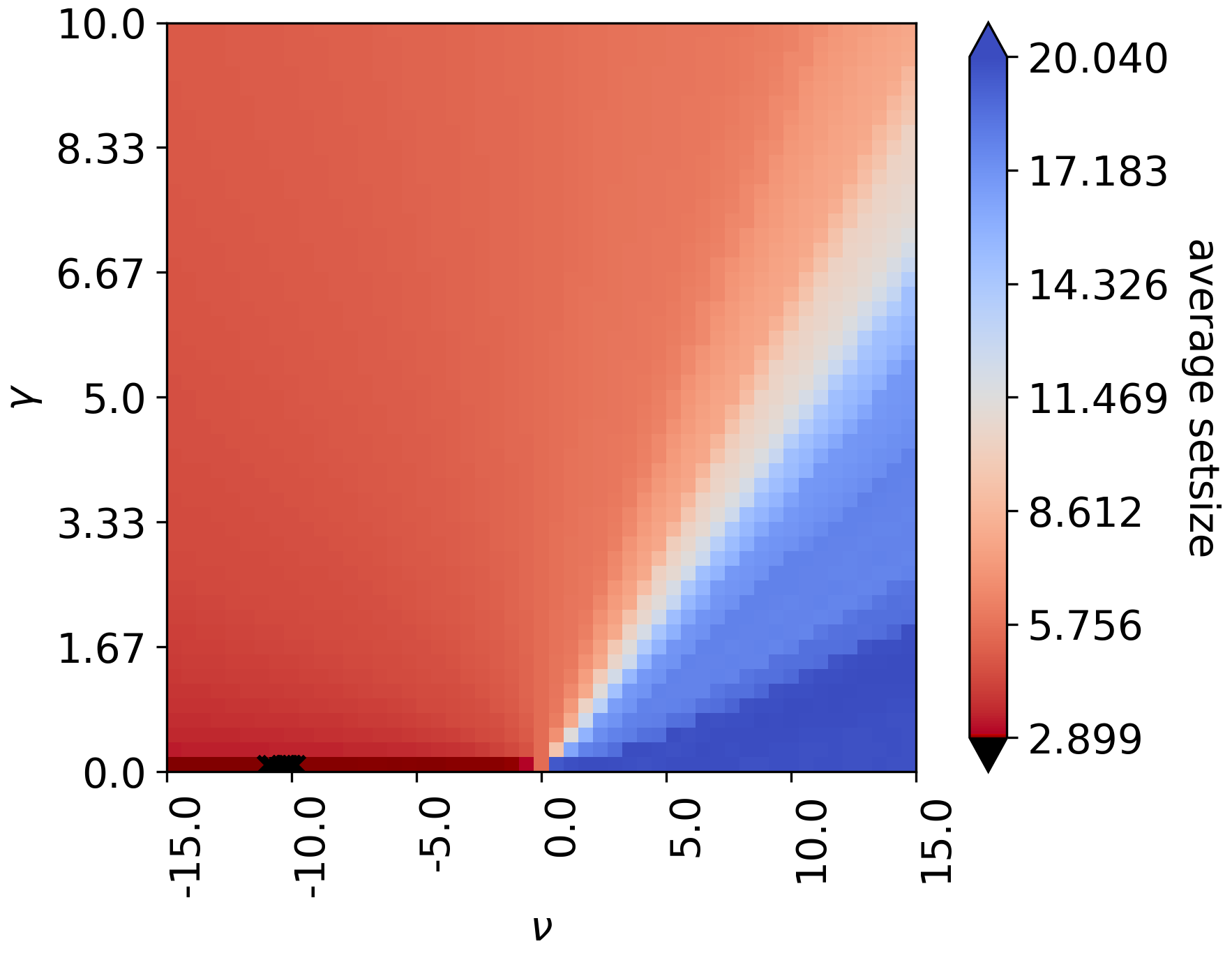} \label{fig:gridd_scipy_av_setsize_margin2}}
    \subfigure[\nonconf{Maxscore2}]{\includegraphics[width=0.235\textwidth]{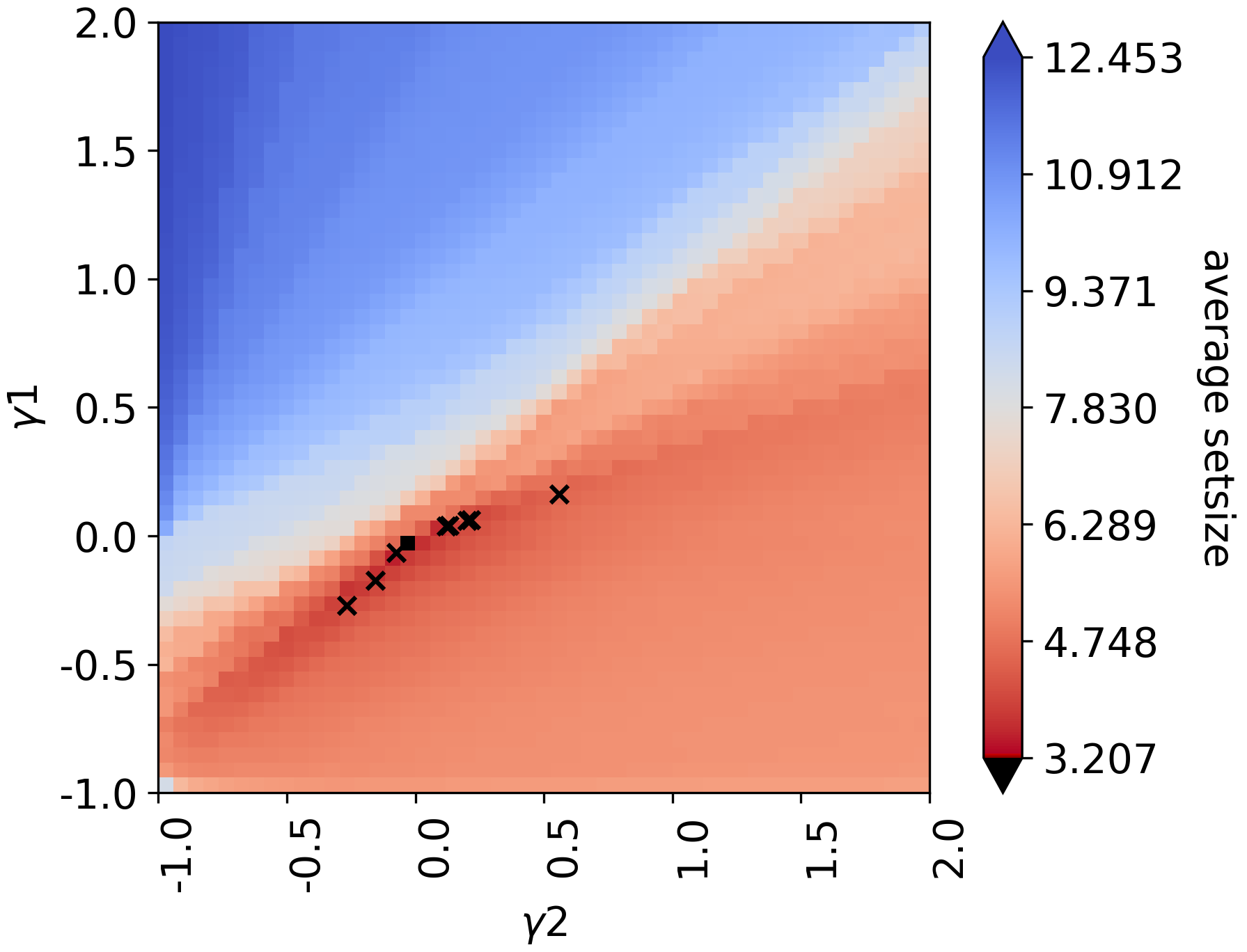} \label{fig:gridd_scipy_av_setsize_maxscore2}}
    \caption{Example grid plots for four different nonconformity measures with Scipy optimisations (black cross), minimising the average set size. \changed{The definition of each nonconformity measure is given in \cref{tab:nonconformity_measures}.}} 
    \label{fig:gridd_scipy_av_setsize}
\end{figure*}

Repeating the optimisation procedure for minimising the average set size, the results are shown in \cref{fig:gridd_scipy_av_setsize}. Similar to the $F_1$ score, we again find that the \nonconf{entropy}, \nonconf{entropy softmax} and \nonconf{maxscore2} nonconformity measures optimise to the \nonconf{baseline} measure and that the \nonconf{entropy} and \nonconf{entropy softmax} measures are degenerate along the zero-axes when no regularisation is used. The \nonconf{margin2} measure optimises to $\gamma=0$, $\nu=-11.9\pm3.5$, for which the average set size is slightly larger but comparable to the \nonconf{baseline} measure. 

Although not equivalent, the results from using the $F_1$ score and the average set size metrics are very similar, suggesting that the metrics are strongly correlated.

%% file: nonconf_comparisons.tex
\input{optimising_results_table}

\subsection{Comparison of nonconformity measures} \label{sec:comparison}
Having optimised each nonconformity measure, we can now compare their performance \changed{on the evaluation dataset}. The results from optimising all of our nonconformity measures for the three chosen metrics are summarised in \cref{tab:optimisation_results}. The optimal parameters obtained are the mean values over five optimisation runs with different calibration-test data splits \changed{of the optimisation dataset}. The uncertainties are the standard deviations over these runs. The plots shown and discussed in \cref{sec:optimising_nonconf} each represent \changed{the first} of these five \changed{optimisation} runs. The evaluation scores are calculated using the optimised parameters and the evaluation dataset. The uncertainties in the evaluation scores stem from the uncertainty of the parameter optimisation. 

For the $F_1$ score, as noted in the previous section, most of the nonconformity measures optimise to parameters that reduce them to the baseline. The exceptions are the \nonconf{softmax}, \nonconf{margin2} and \nonconf{Brier} nonconformity measures, which do not reduce to the \nonconf{baseline} measure but achieve $F_1$ scores comparable to the \nonconf{baseline}. 

Using singletons, we find that the baseline is no longer the best and that all other nonconformity measures perform slightly better, with \nonconf{maxscore2} giving the highest number. However, the evaluation scores are all very similar, and the non-zero standard deviations on the parameters indicate that there still is some statistical uncertainty. 

Using the average set size, the behaviour of the nonconformity measures is similar to using $F_1$ scores. All nonconformity measures either reduce to the baseline for optimal parameters or return comparable average set sizes. 

\begin{figure*}
    \centering
    \subfigure[$F_1$ score]{\includegraphics[width=0.32\textwidth]{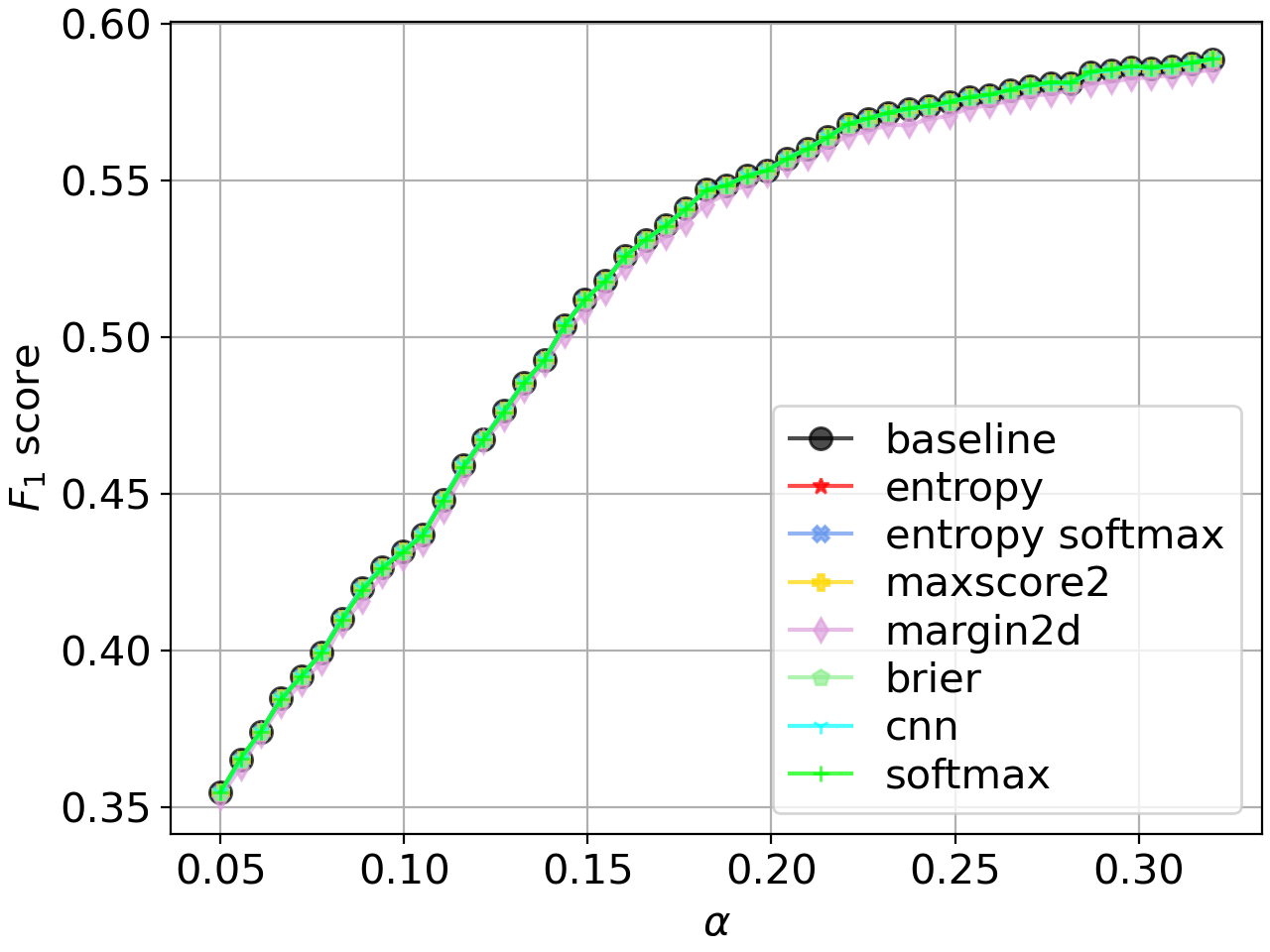} \label{fig:nonconf_comparison_f1}}
    \subfigure[Singletons]{\includegraphics[width=0.32\textwidth]{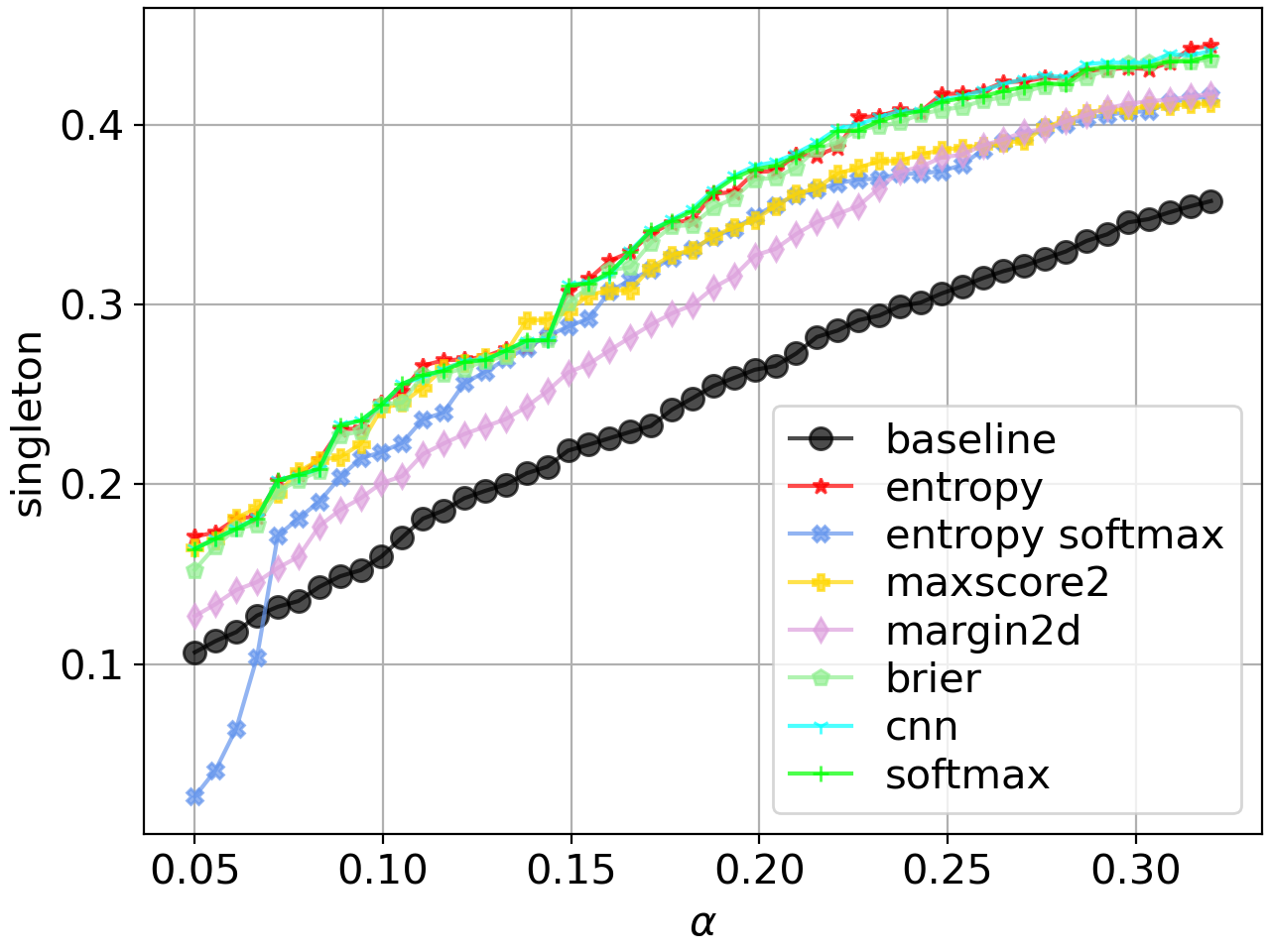}}
    \subfigure[Average set size]{\includegraphics[width=0.32\textwidth]{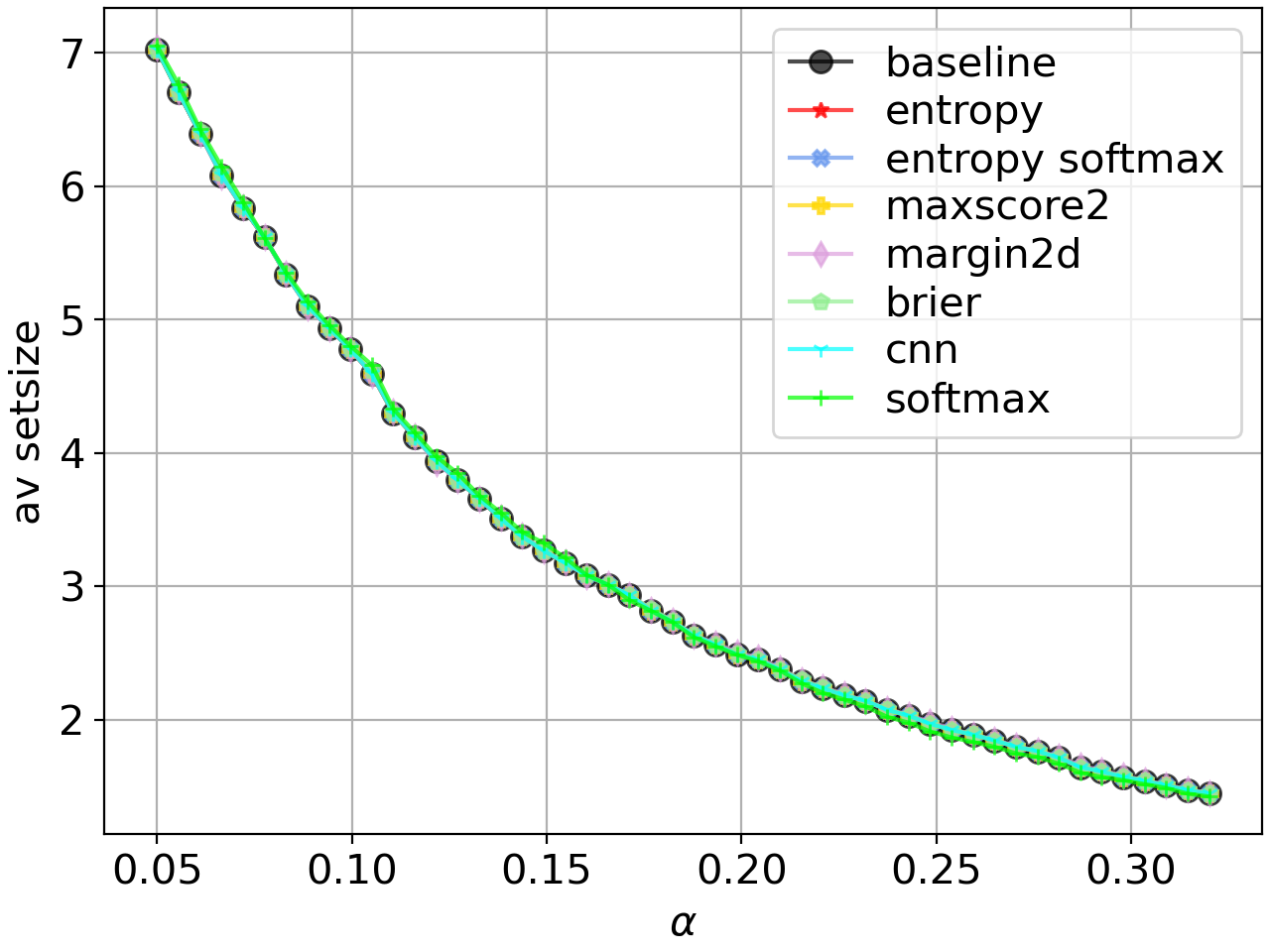} \label{fig:nonconf_comparison_av_setsize}}
    \caption{Comparison of all the discussed nonconformity measures, making use of different metrics. \label{fig:nonconf_comparison_singleton}} 
    \label{fig:nonconf_comparisons}
\end{figure*}

To explore if the optimisation depends on the error rate $\alpha$, we run repeated tests for varying $\alpha$. The plots in \cref{fig:nonconf_comparisons} show the optimised nonconformity measures together with the \nonconf{baseline} over varying error rates $\alpha$. We note that, as expected, the results from all metrics vary with varying $\alpha$. Furthermore, the relative performance of the different nonconformity measures depends on the value of $\alpha$ \changed{for the singleton metric only}. 

For the $F_1$ score and average set size metrics, all optimised nonconformity measures either reduce to or are comparable to the baseline measure. This is true regardless of the value of $\alpha$, as shown in \cref{fig:nonconf_comparison_f1} and \cref{fig:nonconf_comparison_av_setsize}. The optimal parameters for each nonconformity measure were found for $\alpha=0.1$ and these optimised parameters were then applied for other values of $\alpha$. However, a brief investigation shows that optimising at other values of $\alpha$ does not change the results significantly and that the baseline measure is still preferred. 

As shown in both \cref{fig:nonconf_comparison_singleton} and \cref{tab:optimisation_results}, all optimised nonconformity measures considered return a higher average number of singletons than the baseline measures for all $\alpha$. For this metric, the relative performance of nonconformity measures varies slightly with varying $\alpha$, and most notably does the \nonconf{entropy softmax} measure perform significantly worse at lower $\alpha$. 

Our results thus show that the same nonconformity measures optimise and perform differently when considering different metrics. Hence, when applying \ac{CP} to a new problem it is worth considering not only which nonconformity measure will perform best but also which metric we are most interested in. For example for our application of \ac{CP} to Gravity Spy; if we value overall smaller uncertainties we should minimise the average set size to find our optimal nonconformity measure, but if we value uniquely classifying the glitches we should maximise the number of singletons. The $F_1$ score is a metric considering various aspects of the performance of \ac{CP}. For our application, we found that the $F_1$ score results are similar to the average set size and thus the size of the prediction sets appears to have a greater impact than how often glitches are uniquely classified correctly. From the definition of the $F_1$ score in \cref{eq:f1}, it is evident that the number of incorrect classifications (\ac{FP} and \ac{FN}) increases with larger prediction sets. Therefore, we see that the $F_1$ score and average set size metric will be strongly correlated, explaining the similarity of results when used as a loss function.

%% file: optimising_results_table.tex
\begin{table*}
  \caption{Optimisation results at $\alpha=0.1$. To estimate uncertainties, we calculate the standard deviations in the fitted parameters over multiple runs on different data splits. However, in some cases, all runs produce identical results for the fitted parameters; in these cases, we instead take a conservative estimate of the uncertainty by giving the bin size \changed{from gridding}.}
  \label{tab:optimisation_results}
  \centering
  \begin{ruledtabular}
  \begin{tabular}{lllllll}
    Nonconformity measure & \multicolumn{3}{c}{Optimised parameters} & \multicolumn{3}{c}{Evaluation scores}  \\
    \cmidrule(r){2-4} \cmidrule(r){5-7}
     & $F_1$ score & singletons & set size & $F_1$ score & singletons & set size \\
    \midrule
    
    Baseline & - & - & - & 0.456 & 0.179 & 4.263   \\
    
    Softmax & $\beta=0.001\pm0.001$ & $\beta=0.51\pm0.39$ & $\beta=0.001\pm0.001$ &  $0.455\pm10^{-4}$ & $0.261\pm0.006$ & $4.324\pm0.026$  \\

    Entropy & $\gamma=0\pm0.01$ & $\gamma=0.14\pm0.05$ & $\gamma=0\pm0.01$ & $0.456\pm0.003$ & $0.267\pm0.003$ &  $4.263\pm0.04$  \\ 
     & $\nu=0\pm0.01$  & $\nu=-0.2\pm0.01$ & $\nu=0\pm0.01$ &  &  &   \\ 
    
    Entropy softmax & $\nu=0\pm0.01$ & $\nu=1.8\pm0.2$  & $\nu=0\pm0.01$ & $0.456\pm10^{-6}$ & $0.237\pm0.007$ & $4.263\pm10^{-4}$   \\
     & $\beta=0\pm0.01$ & $\beta=-6.5\pm0.4$ & $\beta=0\pm0.01$ & & &   \\

    Maxscore2 &  $\gamma_1=0\pm0.01$ &  $\gamma_1=-0.4\pm0.2$ & $\gamma_1=0\pm0.01$  & $0.456\pm0.006$ & $0.271\pm0.002$ & $4.263\pm0.074$  \\ 
     &  $\gamma_2=0\pm0.01$ &  $\gamma_2=-0.4\pm0.2$ & $\gamma_2=0\pm0.01$  & & &  \\ 
    
    Margin2 & $\gamma=0\pm0.01$ & $\gamma=0.2\pm0.01$ & $\gamma=0\pm0.01$ & $0.455\pm10^{-6}$ & $0.264\pm0.002$ & $4.276\pm0.002$   \\
     & $\nu=-8.8\pm1.5$ & $\nu=-7.4\pm3.8$ &  $\nu=-11.9\pm3.5$ & & &   \\
     
    CNN & $\gamma=1.0\pm0.001$ & $\gamma=0.96\pm0.04$ & $\gamma=1.0\pm0.001$ & $0.456\pm0.006$ & $0.266\pm0.006$ & $4.263\pm0.075$  \\

    Brier & $\nu=1.001\pm0.002$ & $\nu=1.012\pm0.015$  & $\nu=1.002\pm0.001$ & $0.448\pm0.005$ & $0.263\pm0.006$ & $4.671\pm0.011$ \\
    
  \end{tabular}
  \end{ruledtabular}
\end{table*}

%% file: optimising_indevidual.tex
\subsection{Individual glitch classes}
It is important to note that the results in the previous section are calculated for Gravity Spy and the ensemble of all glitches in our data set and that other algorithms or other datasets for the same algorithm may perform differently. 

\begin{figure*}
    \centering
    \subfigure[\nonconf{Entropy}, Koi Fish]{\includegraphics[width=0.235\textwidth]{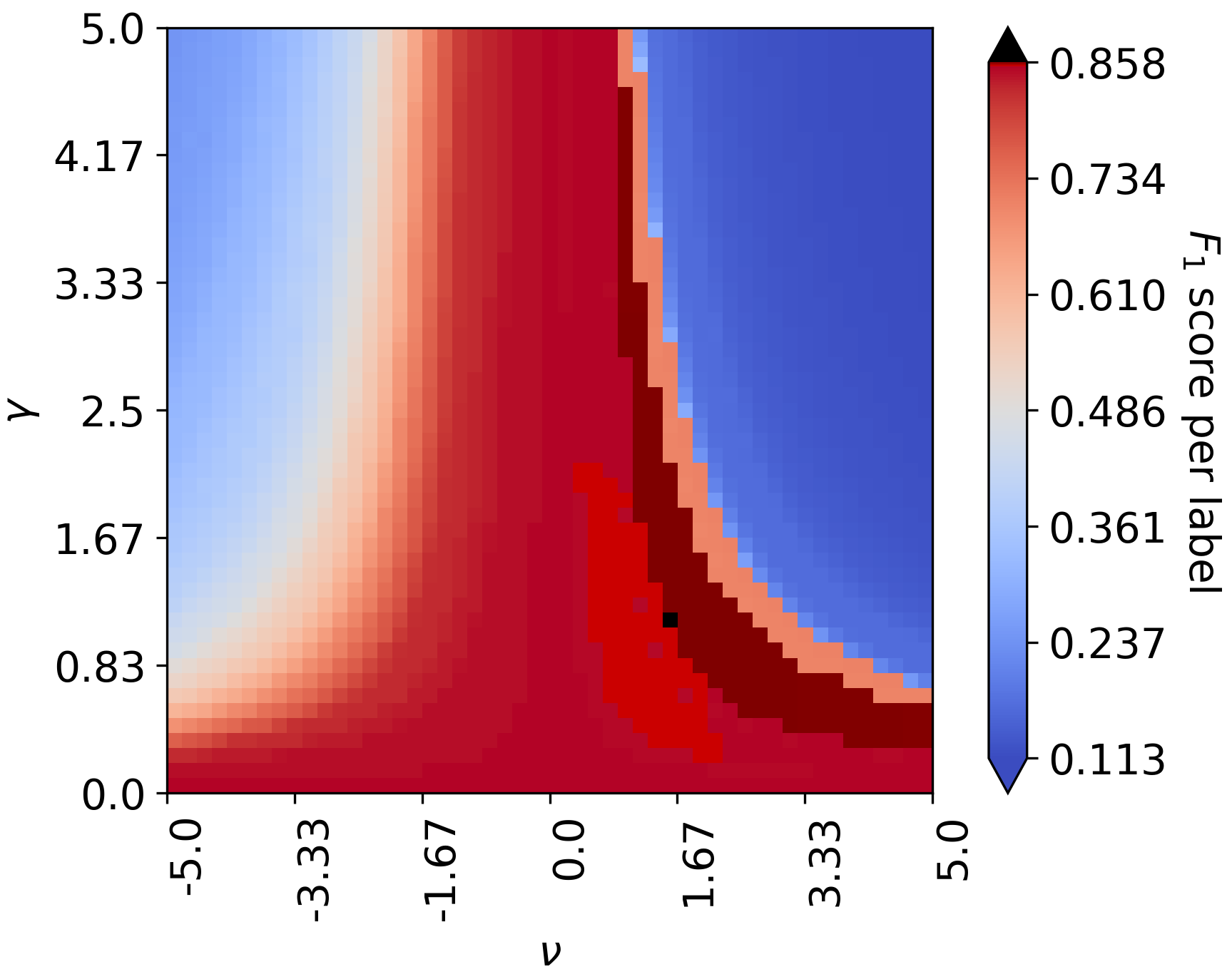} \label{fig:gridd_scipy_individual_entropy_koi_fish}}
    \subfigure[\nonconf{Entropy softmax}, Low Frequency Burst]{\includegraphics[width=0.235\textwidth]{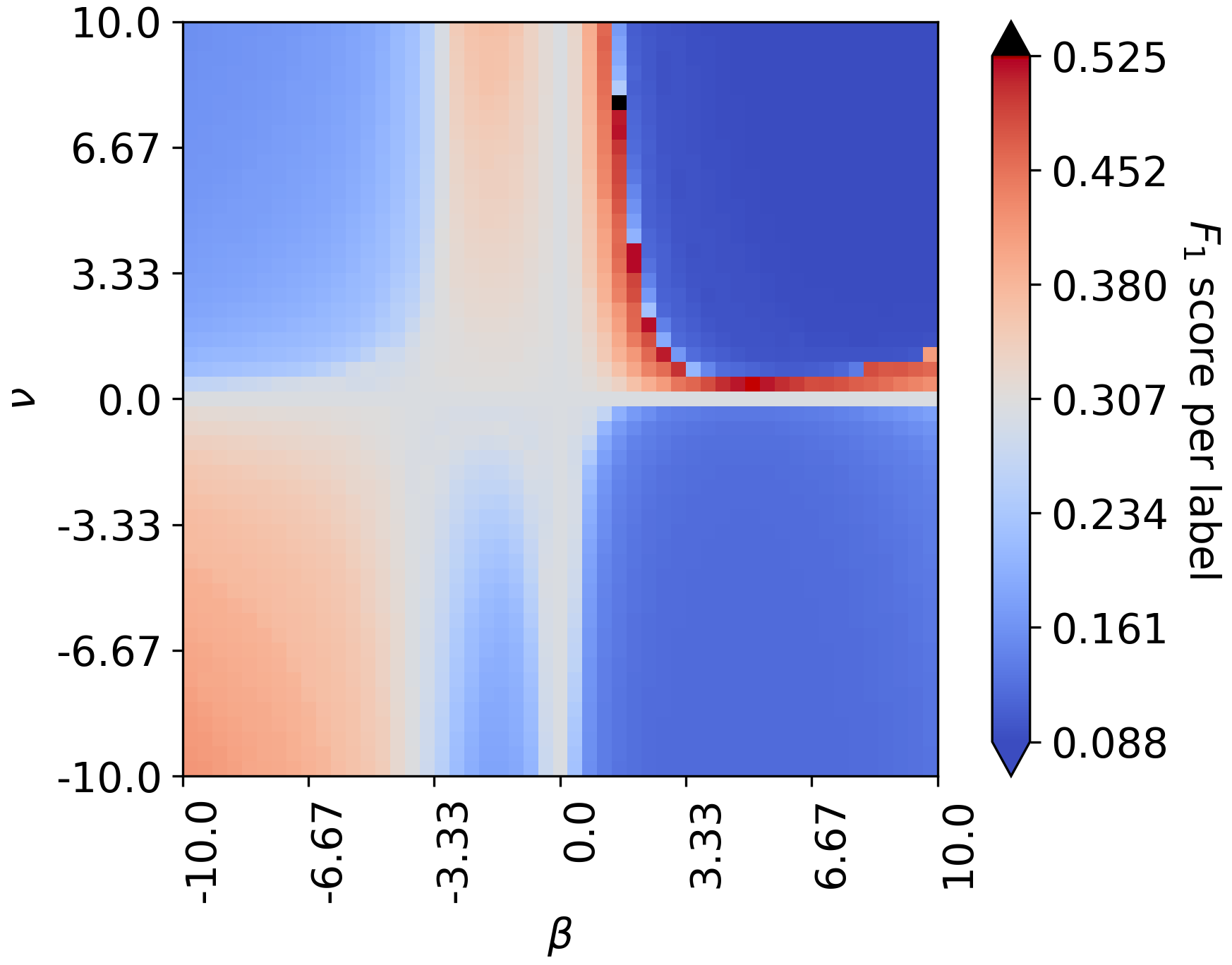} \label{fig:gridd_scipy_individual_entropy_softmax_low_freq_burst}}
    \subfigure[\nonconf{Maxscore2}, Low Frequency Burst]{\includegraphics[width=0.235\textwidth]{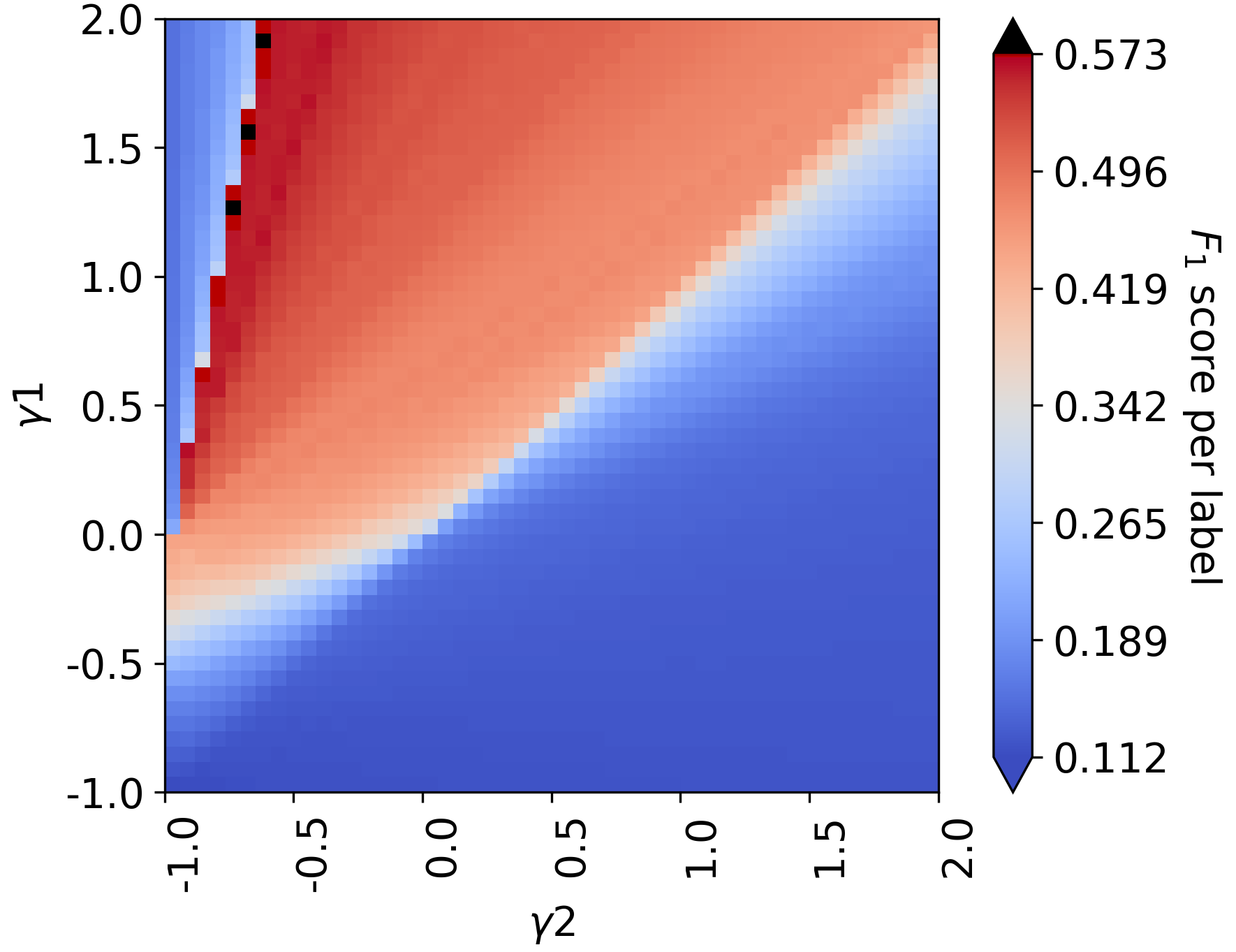} \label{fig:gridd_scipy_individual_maxscore2_low_freq_burst}} 
    \subfigure[\nonconf{Maxscore2}, Wandering Line]{\includegraphics[width=0.235\textwidth]{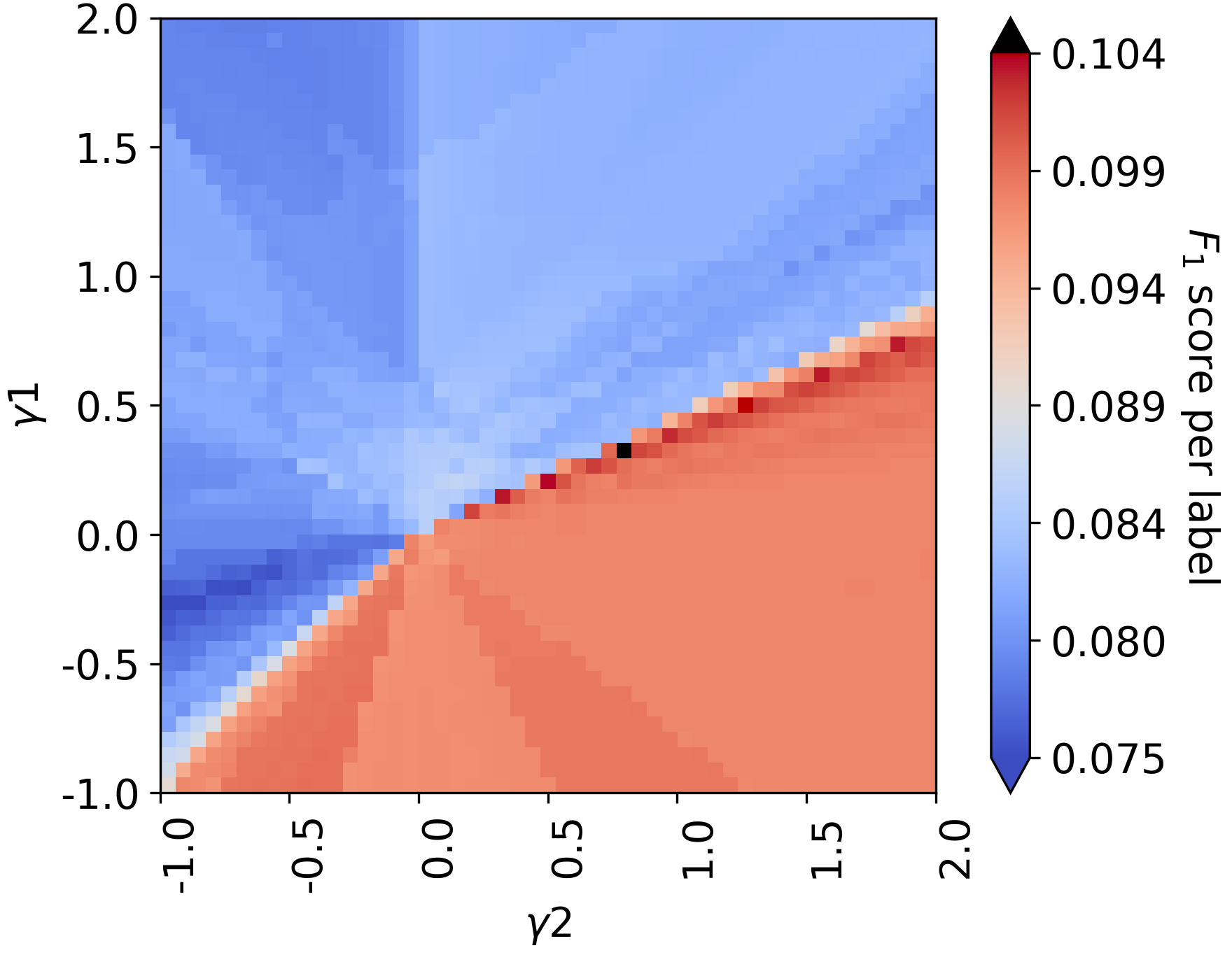} \label{fig:gridd_scipy_individual_maxscore2_wandering_line}} 
    \caption{Example grid plots of individual glitch classes using the $F_1$ score. The plots show that in these cases the nonconformity measures do not optimise to the \nonconf{baseline}.} 
    \label{fig:gridd_scipy_individual}
\end{figure*}

As an example, we can treat each glitch class as an individual dataset and perform the same optimisation. We find that for most of the glitch classes, the results agree with the results in section \cref{sec:optimising_nonconf}, as expected. However, for some of the classes, other nonconformity measures return higher $F_1$ scores than the \nonconf{baseline}. In \cref{fig:gridd_scipy_individual}, a few example plots where the $F_1$ score does not optimise to the \nonconf{baseline} are shown. 

We further note that the same nonconformity measure can behave very differently for different glitch datasets, as seen for example in comparing the plots in \cref{fig:gridd_scipy_individual_maxscore2_low_freq_burst} and \cref{fig:gridd_scipy_individual_maxscore2_wandering_line}. 

This demonstrates how the choice of nonconformity measure is problem-specific. Applying the analysis from this section to another dataset or problem setup may give different results. 

\begin{figure*}
    \centering
    \subfigure[$F_1$ score]{\includegraphics[width=0.325\textwidth]{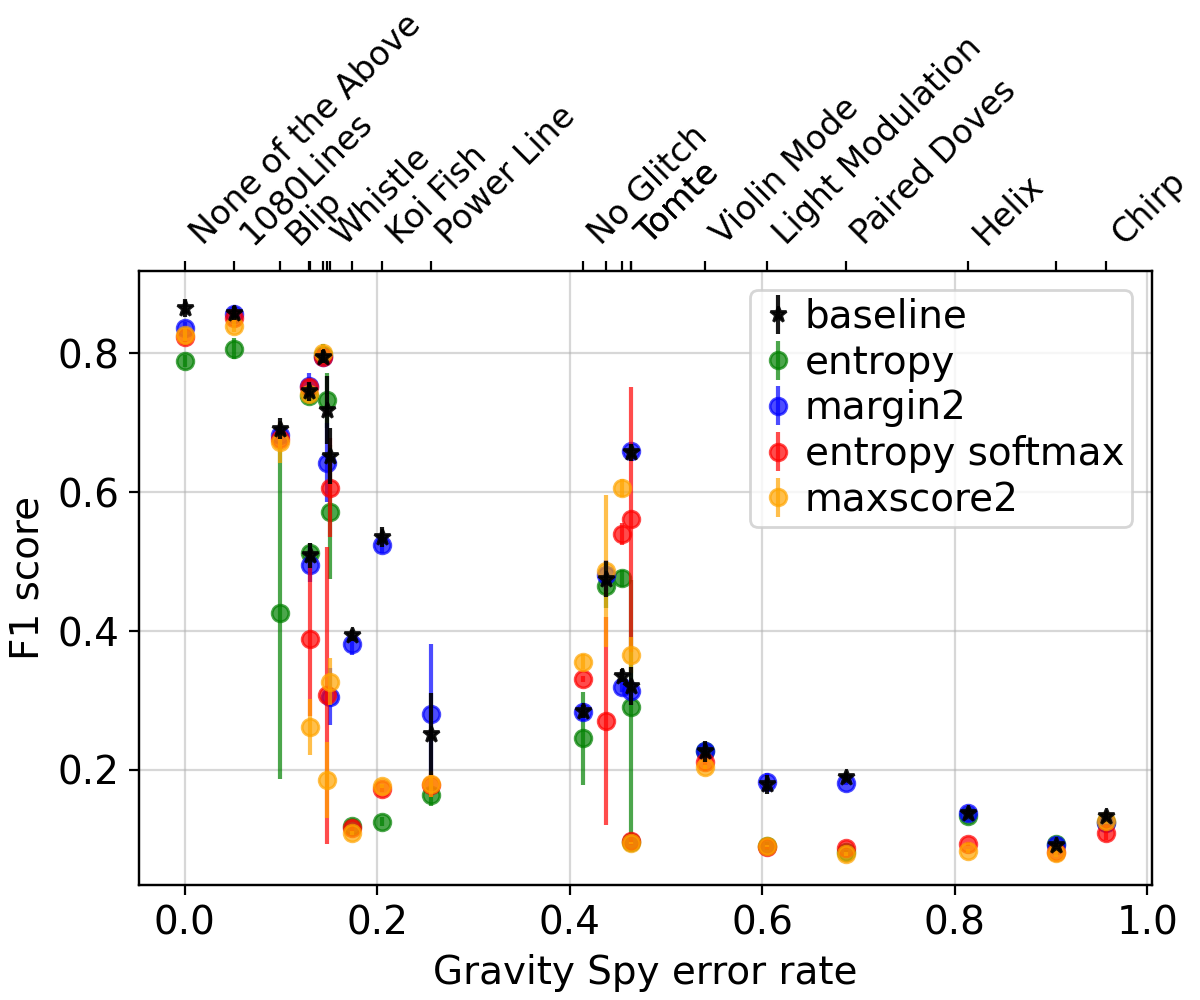} \label{fig:error_vs_f1}}
    \subfigure[Singletons]{\includegraphics[width=0.325\textwidth]{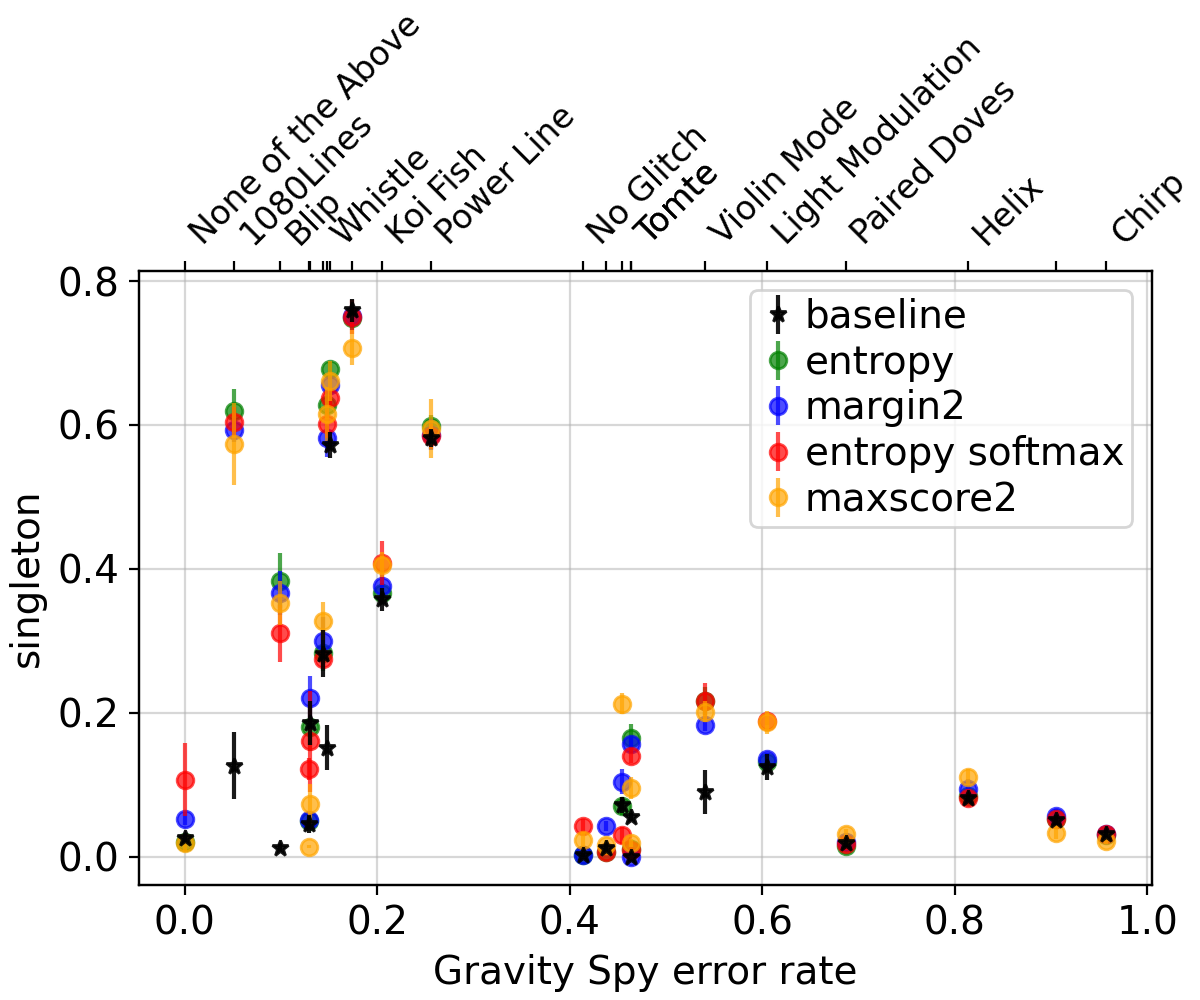}\label{fig:error_vs_singleton}}
    \subfigure[Average set size]{\includegraphics[width=0.325\textwidth]{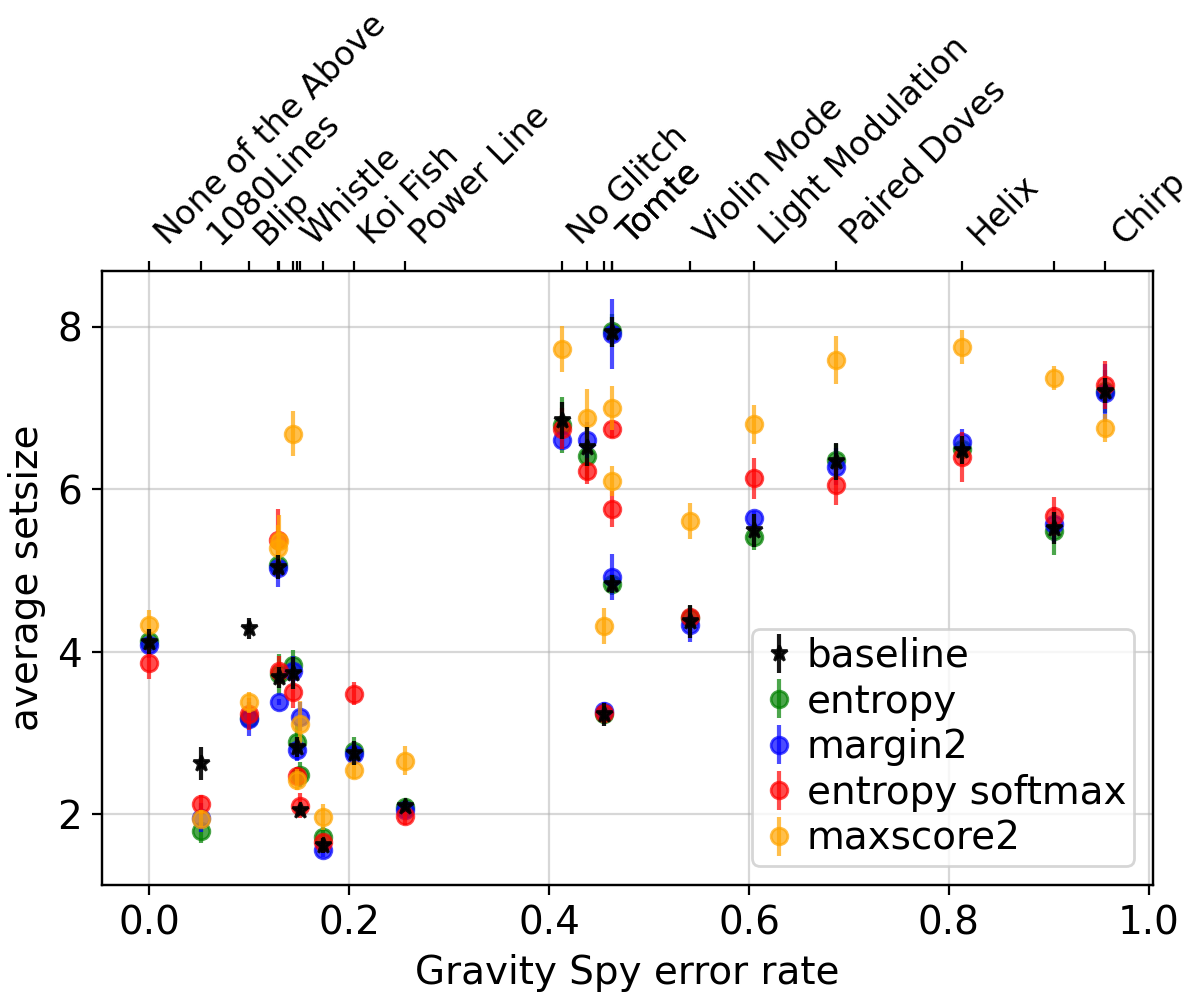} \label{fig:error_vs_avset}}
    \caption{\changed{The plots show the} Gravity Spy (GS) error rate of individual glitch class datasets versus label-specific scores for all metrics for a few of the discussed nonconformity measures. \changed{Each error-rate point on the $x$-axis corresponds to one glitch class, and a few of the classes are named in the top axis (we do not show all for clarity). The nonconformity measures} have been optimised for each glitch class individually. The error bars represent the standard deviation obtained from repeating the evaluation calculation for different splits of each dataset.} 
    \label{fig:error_vs_score}
\end{figure*}

The different glitch classes have varying Gravity Spy classification accuracy, with some classes being classified correctly more often than others. To investigate how the performance changes when applying \ac{CP} to the individual glitch datasets, we optimise our nonconformity measures for each of these \changed{glitch-specific} datasets individually. We can then calculate evaluation sores for each optimised nonconformity measure and glitch-specific dataset, and plot them against the error rate for the respective glitch class, see \cref{fig:error_vs_score}. We find that, a higher classification accuracy (lower Gravity Spy error rate) for a glitch class implies that that dataset will generally achieve higher overall $F_1$ scores, see \cref{fig:error_vs_f1}, and smaller average prediction set sizes, see \cref{fig:error_vs_avset}. The correlation for the singleton metric is less clear, as shown in \cref{fig:error_vs_singleton}. 

Furthermore, the plots in \cref{fig:error_vs_score} confirm the previous discussion that different metrics are preferred for different glitch datasets. However, there does not seem to be any correlation between the classification accuracy of a glitch class and which nonconformity measure is preferred for that dataset. 

%% file: discussion.tex
In this work, we have demonstrated the application of \ac{CP} to the Gravity Spy glitch classification \ac{ML} algorithm. We have discussed properties of \ac{CP} such as its guaranteed validity and particularly focused on the nonconformity measure. 

The nonconformity measure is a key element of \ac{CP}, which transforms the heuristic output of the underlying algorithm into a rigorous uncertainty. Since the optimal choice of nonconformity measure is not theoretically predicted, we parameterise families of possible measures and then investigate the choice of three metrics, the $F_1$ score, average prediction set size, and number of singletons, in optimising the nonconformity parameters. We have applied this to our Gravity Spy test case, but the methodology could be applied in general to identify optimal nonconformity measures.

For the $F_1$ score and average set seize metrics, the results show that the simple, and most common, \nonconf{baseline} nonconformity measure returns the best scores and that, even after optimising, none of the other nonconformity measures are better. In fact, all nonconformity measures which reduce to the \nonconf{baseline} measure for certain parameter values are optimised for these parameters. Meanwhile, all optimised nonconformity measures we have considered return a higher number of singletons than the \nonconf{baseline} measure. Hence, our results show that the choice of nonconformity measure should depend on the metric of interest. For example, if the overall certainty of predictions (small prediction set sizes) is important, we should choose the \nonconf{baseline} measure. However, if creating a set of correctly classified glitches of a certain class is the aim (maximised singletons), the \nonconf{maxscore2} measure will return the best result for our Gravity Spy test application.  

Furthermore, considering the $F_1$ scores for individual glitch class datasets, the \nonconf{baseline} measure is not always the best, and some of the nonconformity measures optimise to parameters that do not reduce them to the \nonconf{baseline}. We can thus conclude that choosing the optimal nonconformity measure is complicated and depends both on the algorithm \ac{CP} is applied to as well as the dataset used and the metric of interest. 

\changed{Knowing that different nonconformity measures are preferred for the different glitch class datasets, one option to improve performance for the full dataset further would be to mix different measures for different classes, such that each class uses the preferred nonconformity measure. Alternatively, adopting the measure that works best overall is a solution. As we have shown in this work, the \nonconf{baseline} measure gives the best results for the $F_1$ score for the overall dataset and is also most often preferred for the class-specific datasets for this metric. This confirms the intuition that if a nonconformity score is good for the majority of smaller datasets, it will also be good for the full dataset.}

Comparing our results for the \nonconf{baseline}, \nonconf{CNN} and \nonconf{Brier} nonconformity measures to similar work in the literature, we confirm that the optimal nonconformity measure is dependent on the application. In Ref. \cite{matiz2019inductiveCPforNN}, the authors found that the \nonconf{CNN} nonconformity measure returned the highest number of singletons for $\gamma=0$ while the smallest average set size was found for $\gamma=1$ for their application of a \ac{CNN} to face and object recognition databases. For our Gravity Spy application, we also found that the \nonconf{CNN} nonconformity measure is optimised for $\gamma=1$ for the average set size metric; however, for the singleton metric, we find $\gamma=0.95$, which differs from the result in \cite{matiz2019inductiveCPforNN}. Similarly to our results, the authors in Ref. \cite{johansson2017modelagnosticNonconf} found that the \nonconf{baseline} nonconformity measure produces a smaller average set size than the \nonconf{brier} measure, while the \nonconf{brier} measures returned a higher number of singletons than the \nonconf{baseline} for their application, confirming again that the optimal choice of nonconformity measure depends on the preferred metric. 

To address the significance of our results, we first observe that the optimisation we apply improves the individual performance of all our nonconformity measures for all metrics, as seen from the plots in \cref{sec:optimising_nonconf}. Secondly, the differences between the evaluation scores for the optimised nonconformity measures are small for all metrics, and one could thus argue that choosing one above another would only minimally affect the outcome. Hence, our results on the full Gravity Spy dataset could be taken to justify the use of the \nonconf{baseline} measure for all metrics. However, investigating the example plots for individual glitch class datasets in \cref{fig:gridd_scipy_individual}, shows that the difference in $F_1$ scores between where the measures are optimised and where they reduce to the \nonconf{baseline} are no longer as small (for example, the difference in $F_1\in[0,1]$ score is $\approx0.2$ for \cref{fig:gridd_scipy_individual_entropy_softmax_low_freq_burst}). In these cases, other nonconformity measures return notably better results and using the \nonconf{baseline} would be less justified.

While the optimised values found in this paper are only applicable to the Gravity Spy glitch classification algorithm and the dataset we have used, the methods we have described in this paper are applicable to any point-prediction algorithm. \ac{CP} can be used to add uncertainty to classification algorithms in the manner described in this paper, but can also be used to add confidence intervals to \ac{ML} regression algorithms \cite{shafer2008tutorialonCP} where it is otherwise difficult to obtain uncertainty quantifications. The method for the regression case is almost identical to the classification case, with the main difference being a different choice of nonconformity measure. 

While uncertainties are important in themselves, \ac{CP} can also be used to compare algorithm performances quantitatively, or optimally combine the output from multiple algorithms. For example, \ac{CP} together with the quantification metrics discussed in this paper could be used to investigate which algorithm performs a given task better or if a combination of algorithms gives the optimal result. \changed{By defining a metric based on the prediction sets (such as those discussed in this paper), \ac{CP} can be applied to each algorithm and the most optimal one for the given task and chosen metric can be determined (similarly to how we have found optimal nonconformity measures in this work).} It is also possible to use \ac{CP} to improve the underlying algorithm itself \cite{stutz2021conformaltraining}. 

To continue this work, \ac{CP} could be built into or around the Gravity Spy algorithm for future analysis, providing point predictions with uncertainties. This idea also extends to other gravitational wave applications, for example, searches for compact binary coalescence signals \cite{ashton2024CP}, or could be applied to newly evolving \ac{ML} methods \cite{skliris2020MLy, marx2024Aframe}. \ac{CP} can be used to combine search pipelines to improve performance or provide uncertainties for a \ac{ML} based parameter estimation algorithm. In summary, there are a multitude of applications, for gravitational wave science and otherwise, where \ac{CP} can be useful.

The code for \cref{sec:cp} is openly available from the Zenodo repository \cite{paper_code}.